\documentclass{eptcs-modified}
 \providecommand{\event}{} 
 \usepackage{breakurl}             
\usepackage{amsmath}
\usepackage{amssymb}
\usepackage{amsthm}
\usepackage{tikz}
\usepackage{amscd}
\usetikzlibrary{arrows,chains,matrix,positioning,scopes}
\makeatletter
\tikzset{join/.code=\tikzset{after node path={%
\ifx\tikzchainprevious\pgfutil@empty\else(\tikzchainprevious)%
edge[every join]#1(\tikzchaincurrent)\fi}}}
\makeatother
\tikzset{>=stealth',every on chain/.append style={join},
         every join/.style={->}}
\tikzstyle{labeled}=[execute at begin node=$\scriptstyle,
   execute at end node=$]

\title{Computability on the space of countable ordinals}
\author{
Arno Pauly
\institute{Universit\'e Libre de Bruxelles, Brussels, Belgium\footnote{This work was begun while the author was at the University of Cambridge, UK.}}
\email{Arno.Pauly@cl.cam.ac.uk}
}
\def\titlerunning{Computability on the countable ordinals}
\def\authorrunning{A. Pauly}

\begin{document}
\theoremstyle{definition}
\newtheorem{theorem}{Theorem}
\newtheorem{definition}[theorem]{Definition}
\newtheorem{problem}[theorem]{Problem}
\newtheorem{assumption}[theorem]{Assumption}
\newtheorem{corollary}[theorem]{Corollary}
\newtheorem{proposition}[theorem]{Proposition}
\newtheorem{lemma}[theorem]{Lemma}
\newtheorem{observation}[theorem]{Observation}
\newtheorem{fact}[theorem]{Fact}
\newtheorem{question}[theorem]{Open Question}
\newtheorem{conjecture}[theorem]{Conjecture}
\newtheorem{example}[theorem]{Example}
\newtheorem*{theorem*}{Theorem}
\newcommand{\dom}{\operatorname{dom}}
\newcommand{\id}{\textnormal{id}}
\newcommand{\Cantor}{{\{0, 1\}^\mathbb{N}}}
\newcommand{\Baire}{{\mathbb{N}^\mathbb{N}}}
\newcommand{\Lev}{\textnormal{Lev}}
\newcommand{\hide}[1]{}
\newcommand{\mto}{\rightrightarrows}
\newcommand{\uint}{{[0, 1]}}
\newcommand{\bft}{\mathrm{BFT}}
\newcommand{\lbft}{\textnormal{Linear-}\mathrm{BFT}}
\newcommand{\pbft}{\textnormal{Poly-}\mathrm{BFT}}
\newcommand{\sbft}{\textnormal{Smooth-}\mathrm{BFT}}
\newcommand{\ivt}{\mathrm{IVT}}
\newcommand{\cc}{\textrm{CC}}
\newcommand{\lpo}{\textrm{LPO}}
\newcommand{\llpo}{\textrm{LLPO}}
\newcommand{\aou}{AoU}
\newcommand{\Ctwo}{C_{\{0, 1\}}}
\newcommand{\name}[1]{\textsc{#1}}
\newcommand{\C}{\textrm{C}}
\newcommand{\UC}{\textrm{UC}}
\newcommand{\ic}[1]{\textrm{C}_{\sharp #1}}
\newcommand{\xc}[1]{\textrm{XC}_{#1}}
\newcommand{\me}{\name{P}.~}
\newcommand{\etal}{et al.~}
\newcommand{\eval}{\operatorname{eval}}
\newcommand{\rank}{\operatorname{rank}}
\newcommand{\Sierp}{Sierpi\'nski }
\newcommand{\isempty}{\operatorname{IsEmpty}}
\newcommand{\spec}{\textrm{Spec}}
\newcommand{\cord}{\textrm{COrd}}
\newcommand{\Cord}{\textrm{\bf COrd}}
\newcommand{\CordM}{\Cord_{\textrm{M}}}
\newcommand{\CordK}{\Cord_{\textrm{K}}}
\newcommand{\CordHL}{\Cord_{\textrm{HL}}}
\newcommand{\leqW}{\leq_{\textrm{W}}}
\newcommand{\leW}{<_{\textrm{W}}}
\newcommand{\equivW}{\equiv_{\textrm{W}}}
\newcommand{\geqW}{\geq_{\textrm{W}}}
\newcommand{\pipeW}{|_{\textrm{W}}}
\newcommand{\Det}{\textrm{Det}}
\newcommand{\bcode}{\textrm{BC}}

\newcommand\tboldsymbol[1]{%
\protect\raisebox{0pt}[0pt][0pt]{%
$\underset{\widetilde{}}{\boldsymbol{#1}}$}\mbox{\hskip 1pt}}

\newcommand{\bolds}{\tboldsymbol{\Sigma}}
\newcommand{\boldp}{\tboldsymbol{\Pi}}
\newcommand{\boldd}{\tboldsymbol{\Delta}}
\newcommand{\boldg}{\tboldsymbol{\Gamma}}

\newcounter{saveenumi}
\newcommand{\seti}{\setcounter{saveenumi}{\value{enumi}}}
\newcommand{\conti}{\setcounter{enumi}{\value{saveenumi}}}

\maketitle

\begin{abstract}
While there is a well-established notion of what a computable ordinal is, the question which functions on the countable ordinals ought to be computable has received less attention so far. We propose a notion of computability on the space of countable ordinals via a representation in the sense of computable analysis. The computability structure is characterized by the computability of four specific operations, and we prove further relevant operations to be computable. Some alternative approaches are discussed, too.

As an application in effective descriptive set theory, we can then state and prove computable uniform versions of the Lusin separation theorem and the Hausdorff-Kuratowski theorem. Furthermore, we introduce an operator on the Weihrauch lattice corresponding to iteration of some principle over a countable ordinal.
\end{abstract}


\section{Introduction}
In \name{Turing}'s seminal paper \cite{turing}, he suggested to call a real number \emph{computable} iff its decimal expansion is. However, in the corrections \cite{turingb}, he pointed out that it is better to use the definition that a real number is computable, iff there is a computable sequence of rational intervals collapsing to it (an idea by \name{Brouwer}). Both definitions yield the same class of real numbers -- but the natural notions of what is a computable function on the real numbers that come along with them differ. For example, $x \mapsto 3x$ is only computable regarding the latter, but not the former notion.

We shall show that there is a similar phenomenon regarding the notion of a computable ordinal: While all usual approaches via numberings (e.g.~\cite{kleene3},\cite[Section 2.8]{weihrauch}) yield the same computable points, the induced uniform structure differs. Like multiplication with $3$ for the real numbers, some simple functions such as the maximum of two ordinals fail to be computable w.r.t.~several common representations of the ordinals. We propose our standard representation in Definition \ref{def:naivekleene} and characterize its equivalence class in Theorem \ref{theo:charac}. Further major results on computability of operations on the countable ordinals are provided in Theorems \ref{theo:newsup}, \ref{theo:binarymin} and \ref{theo:cbaireremoval}.

In Sections \ref{sec:otherstructures} and \ref{sec:alternatives} some alternatives to Definition \ref{def:naivekleene} are investigated, and either proven to be equivalent to it, or a reason to reject them as standard computability structures is exhibited.

As an application, in Section \ref{sec:descriptive} we continue a research programme to investigate concepts from descriptive set theory in the very general setting of represented spaces, and in a fashion that produces both classical and effective results simultaneously. A survey of this approach is given in \cite{pauly-overview-arxiv}. One of the first theorems studied in this way is the Jayne-Rogers theorem (\cite{jaynerogers}, simplified proof in \cite{ros}); a computable version holding also in some non-Hausdorff spaces was proven by the author and \name{de Brecht} in \cite{paulydebrecht} using results about Weihrauch reducibility in \cite{paulybrattka}. Our goal here is to state and prove corresponding versions of the Lusin separation theorem and the Hausdorff-Kuratowski theorem.

Finally, in Section \ref{sec:iteration} we employ the space of countable ordinals to introduce an operator on the Weihrauch lattice that captures iterating some principle over some countable ordinal.

An extended abstract containing preliminary versions of some of the results presented here appeared as \cite{pauly-ordinals-mfcs}.

\subsection{Represented spaces}
We shall briefly introduce the notion of a \emph{represented space}, which underlies computable analysis \cite{weihrauchd}. For a more detailed presentation we refer to \cite{pauly-synthetic}. A represented space is a pair $\mathbf{X} = (X, \delta_X)$ of a set $X$ and a partial surjection $\delta_X : \subseteq \Baire \to X$ (the representation). A represented space is called \emph{complete}, iff it has an equivalent representation that is a total function, cf.~\cite{selivanov5}.

A multi-valued function between represented spaces is a multi-valued function between the underlying sets. For $f : \subseteq \mathbf{X} \mto \mathbf{Y}$ and $F : \subseteq \Baire \to \Baire$, we call $F$ a realizer of $f$ (notation $F \vdash f$), iff $\delta_Y(F(p)) \in f(\delta_X(p))$ for all $p \in \dom(f\delta_X)$.
 $$\begin{CD}
\Baire @>F>> \Baire\\
@VV\delta_\mathbf{X}V @VV\delta_\mathbf{Y}V\\
\mathbf{X} @>f>> \mathbf{Y}
\end{CD}$$
A map between represented spaces is called computable (continuous), iff it has a computable (continuous) realizer. Similarly, we call a point $x \in \mathbf{X}$ computable, iff there is some computable $p \in \Baire$ with $\delta_\mathbf{X}(p) = x$. We write $\mathbf{X} \cong \mathbf{Y}$ to denote that $\mathbf{X}$ and $\mathbf{Y}$ are computably isomorphic.

Given two represented spaces $\mathbf{X}$, $\mathbf{Y}$ we obtain a third represented space $\mathcal{C}(\mathbf{X}, \mathbf{Y})$ of functions from $X$ to $Y$ by letting $0^n1p$ be a $[\delta_X \to \delta_Y]$-name for $f$, if the $n$-th Turing machine equipped with the oracle $p$ computes a realizer for $f$. As a consequence of the UTM theorem, $\mathcal{C}(-, -)$ is the exponential in the category of continuous maps between represented spaces, and the evaluation map is even computable (as are the other canonic maps, e.g.~currying).

Based on the function space construction, we can obtain the hyperspaces of open $\mathcal{O}$, closed $\mathcal{A}$, overt $\mathcal{V}$ and compact $\mathcal{K}$ subsets of a given represented space using the ideas of synthetic topology \cite{escardo}.

Let $\Delta : \subseteq \Baire \to \Baire$ be defined on the sequences containing only finitely many $0$s, and let it map those to their tail starting immediately after the last $0$, with each entry reduced by $1$. This is a surjection. Given a represented space $\mathbf{X} = (X, \delta_\mathbf{X})$, we define the represented space $\mathbf{X}^\nabla := (X, \delta_\mathbf{X} \circ \Delta)$. Informally, in this space, finitely many mindchanges are allowed. The operation $^\nabla$ even extends to an endofunctor on the category of represented spaces \cite{ziegler3,paulydebrecht2,pauly-descriptive-lics}.

\subsection{Weihrauch reducibility}
Several of our results are negative, i.e.~show that certain operations are not computable. We prefer to be more precise, and not to merely state failure of computability. Instead, we give lower bounds for Weihrauch reducibility. The reader not interested in distinguishing degrees of non-computability may skip the remainder of the subsection, and in the rest of the paper (with the exception of Subsection \ref{subsec:hk} and Section \ref{sec:iteration}), read any statement involving Weihrauch reducibility ($\leqW$, $\equivW$, $\leW$) as merely indicating the non-computability of the maps involved.

\begin{definition}[Weihrauch reducibility]
\label{def:weihrauch}
Let $f,g$ be multi-valued functions on represented spaces.
Then $f$ is said to be {\em Weihrauch reducible} to $g$, in symbols $f\leqW g$, if there are computable
functions $K,H:\subseteq\Baire\to\Baire$ such that $K\langle \id, GH \rangle \vdash f$ for all $G \vdash g$.
\end{definition}

The relation $\leqW$ is reflexive and transitive. We use $\equivW$ to denote equivalence regarding $\leqW$,
and by $\leW$ we denote strict reducibility. By $\mathfrak{W}$ we refer to the partially ordered set of equivalence classes. As shown in \cite{paulyreducibilitylattice,brattka2}, $\mathfrak{W}$ is a distributive lattice. The algebraic structure on $\mathfrak{W}$ has been investigated in further detail in \cite{paulykojiro,paulybrattka4}.

A prototypic non-computable function is $\lpo : \Baire \to \{0,1\}$ defined via $\lpo(0^\mathbb{N}) = 1$ and $\lpo(p) = 0$ for $p \neq 0^\mathbb{N}$. The degree of this function was already studied by \name{Weihrauch} \cite{weihrauchc}.

A few years ago several authors (\name{Gherardi} and \name{Marcone} \cite{gherardi}, \name{P.} \cite{paulyreducibilitylattice,paulyincomputabilitynashequilibria}, \name{Brattka} and \name{Gherardi} \cite{brattka3}) noticed that Weihrauch reducibility would provide a very interesting setting for a metamathematical inquiry into the computational content of mathematical theorems. The fundamental research programme was outlined in \cite{brattka3}, and the introduction in \cite{hoelzl} may serve as a recent survey.

\section{The represented space $\Cord$}

\begin{definition}
\label{def:naivekleene}
Let $\cord$ denote the set of countable ordinals. We define a representation $\delta_{\textrm{nK}} : \subseteq \Baire \to \cord$ inductively via:
\begin{enumerate}
\item $\delta_{\textrm{nK}}(0p) = 0$
\item $\delta_{\textrm{nK}}(1p) = \delta_{\textrm{nK}}(p) + 1$
\item $\delta_{\textrm{nK}}(2\langle p_0, p_1, p_2, \ldots \rangle) = \sup_{i \in \mathbb{N}} \delta_{\textrm{nK}}(p_i)$.
\end{enumerate}
The represented space $\Cord$ is introduced as $(\cord,\delta_{\textrm{nK}})$.
\end{definition}

The design of the representation $\delta_{\textrm{nK}}$ immediately establishes the following:

\begin{observation}
\label{obs:deltanktrivial}
The following maps are all computable:
\begin{enumerate}
\item $0 \in \Cord$
\item $\mathalpha{+1} : \Cord \to \Cord$
\item $\sup : \mathcal{C}(\mathbb{N}, \Cord) \to \Cord$
\end{enumerate}
\end{observation}

Even more, we find that amongst all representations of $\cord$, $\delta_{\textrm{nK}}$ contains the most information consistent with the computability of the maps above:
\begin{proposition}
\label{prop:nkinitial}
Let $\delta$ be a representation of $\cord$ such that the following maps are all computable:
\begin{enumerate}
\item $0 \in (\cord, \delta)$
\item $\mathalpha{+1} : (\cord, \delta) \to (\cord, \delta)$
\item $\sup : \mathcal{C}(\mathbb{N}, (\cord, \delta)) \to (\cord, \delta)$.
\end{enumerate}
Then $\id : \Cord \to (\cord, \delta)$ is computable.
\begin{proof}
Induction along the definition of $\delta_{\textrm{nK}}$.
\end{proof}
\end{proposition}

We can also characterize $\Cord$ as the space with the least information rendering a certain operation computable. This operation will be to list all smaller ordinals, a task central to some inductive constructions. For this, let $\{\textrm{Skip}\} \uplus \Cord$ (or  $\{\textrm{Skip}\} \uplus (\cord,\delta)$) denote the disjoint union of the singleton set $\{\textrm{Skip}\}$ and the space of countable ordinals. Then $\mathcal{C}(\mathbb{N}, \{\textrm{Skip}\} \uplus \Cord)$ is the space of partial sequences of ordinals.

\begin{definition}
Let $\textrm{Lower} : \cord \mto \mathcal{C}(\mathbb{N},\{\textrm{Skip}\} \uplus \cord)$ by defined via $(\alpha_i)_{i \in \mathbb{N}} \in \textrm{Lower}(\alpha)$ iff $\{\beta \in \cord \mid \beta < \alpha\} = \{\beta \in \cord \mid \exists i \in \mathbb{N} \ \alpha_i = \beta\}$.
\end{definition}

\begin{proposition}
\label{prop:lower}
$\textrm{Lower} : \Cord \mto \mathcal{C}(\mathbb{N},\{\textrm{Skip}\} \uplus \Cord)$ is computable.
\begin{proof}
Let $[\cdot,\cdot] : \mathbf{X} \times \mathcal{C}(\mathbb{N},\mathbf{X}) \to \mathcal{C}(\mathbb{N},\mathbf{X})$ be defined via $[x,p](0) = x$ and $[x,p](n+1) = p(n)$. Let $\langle \ \ \rangle : \mathcal{C}(\mathbb{N},\mathcal{C}(\mathbb{N},\mathbf{X})) \to \mathcal{C}(\mathbb{N},\mathbf{X})$ be some standard pairing function.

Now we define the computation of $\textrm{Lower}$ inductively: $\textrm{Lower}(0) = (n \mapsto \textrm{Skip})$, $\textrm{Lower}(\alpha + 1) = [\alpha,\textrm{Lower}(\alpha)]$ and $\textrm{Lower}(\sup_{i \in \mathbb{N}} \beta_i) = \langle (i \mapsto \textrm{Lower}(\beta_i)) \rangle$. Considering the definition of an ordinal being smaller than some other ordinal together with induction proves that this indeed forms a correct algorithm for $\textrm{Lower}$.
\end{proof}
\end{proposition}

\begin{proposition}
\label{prop:nkfinal}
Let $\delta$ be a representation of $\cord$ such that $\textrm{Lower} : (\cord, \delta) \mto \mathcal{C}(\mathbb{N},\{\textrm{Skip}\} \uplus (\cord, \delta))$ is computable. Then $\id : (\cord, \delta) \to \Cord$ is computable.
\begin{proof}
We define $\id : (\cord, \delta) \to \Cord$ inductively together with $\iota : (\{\textrm{Skip}\} \uplus (\cord,\delta)) \to \Cord$. With $\widehat{\iota} : \mathcal{C}(\mathbb{N}, (\{\textrm{Skip}\} \uplus (\cord,\delta))) \to \mathcal{C}(\mathbb{N},\Cord)$ we denote the pointwise application of $\iota$. Let $\iota(\textrm{Skip}) = 0$ and $\iota(\alpha) = \id(\alpha) + 1$. Now \[\id = \left (\sup : \mathcal{C}(\mathbb{N},\Cord) \to \Cord \right ) \circ \widehat{\iota} \circ \textrm{Lower}\]
\end{proof}
\end{proposition}

We can gather the results obtained so far to characterize precisely the computability structure on $\Cord$. In doing so, we answer a question raised by Vasco Brattka during CCA 2015.

\begin{theorem}
\label{theo:charac}
The following are equivalent for a representation $\delta :\subseteq \Baire \to \cord$:
\begin{enumerate}
\item $\id : (\cord,\delta) \to \Cord$ is a computable isomorphism.
\item All of the following are computable:
\begin{enumerate}
\item $0 \in (\cord, \delta)$
\item $\mathalpha{+1} : (\cord, \delta) \to (\cord, \delta)$
\item $\sup : \mathcal{C}(\mathbb{N}, (\cord, \delta)) \to (\cord, \delta)$.
\item $\textrm{Lower} : (\cord, \delta) \mto \mathcal{C}(\mathbb{N},\{\textrm{Skip}\} \uplus (\cord, \delta))$
\end{enumerate}
\end{enumerate}
\begin{proof}
Observation \ref{obs:deltanktrivial} and Proposition \ref{prop:lower} together yield the implication $(1) \Rightarrow (2)$. For $(2) \Rightarrow (1)$, we combine Proposition \ref{prop:nkinitial} and Proposition \ref{prop:nkfinal}.
\end{proof}
\end{theorem}

\subsection{Computability on the finite ordinals}
While the finite ordinals and the natural numbers are often identified, it should not be taken for granted that restricting a natural representation of the ordinals to the finite ordinals will yield some representation equivalent to the usual representation of the natural numbers. In fact, besides the usual natural numbers $\mathbb{N}$, additional spaces with the natural numbers as underlying set will make an appearance here: There are spaces $\mathbb{N}_<$, $\mathbb{N}_>$ and $\mathbb{N}^\nabla$, where a number $n$ is represented by a non-decreasing, respectively non-increasing, respectively arbitrary sequence of integers which eventually converge to $n$. For $\mathbb{N}_<$ and $\mathbb{N}_>$ we shall also consider the variations $\overline{\mathbb{N}}_<$ and $\overline{\mathbb{N}}_>$, where we adjoin an element $\infty$: In $\overline{\mathbb{N}}_<$, is represented by any unbounded non-decreasing sequence, in $\overline{\mathbb{N}}_>$ by a sequence containing only a placeholder $\bot$ (and we allow sequences starting with any number of $\bot$, and then ending as a non-increasing sequence of natural numbers).

\begin{observation}
\label{obs:differentnaturals}
$\left (\id : \mathbb{N}^\nabla \to \mathbb{N} \right ) \equivW \left (\id : \mathbb{N}_< \to \mathbb{N} \right ) \equivW \C_\mathbb{N}$; $\left (\id : \mathbb{N}_> \to \mathbb{N} \right ) \equivW \lpo^*$ and $\lpo \leqW \left (\id : \mathbb{N}_> \to \mathbb{N}_< \right )$.
\end{observation}

\begin{observation}
\label{obs:finiteordinals}
$\id : \mathbb{N}_< \to \Cord$ is a computable embedding.
\end{observation}

Next, we shall characterize the open and compact subsets of $\Cord$. For this we need the spaces $\overline{\mathbb{N}}_<$ and $\overline{\mathbb{N}}_>$. We will understand $\alpha < \infty$ for all countable ordinals $\alpha$.

\begin{proposition}
The map $n \mapsto \{\alpha \in \cord \mid \alpha \geq n\} : \overline{\mathbb{N}}_> \to \mathcal{O}(\Cord)$ is a computable isomorphism.
\begin{proof}
First we show that the map is computable. Given a natural number $n \in \mathbb{N}$ and an ordinal $\alpha \in \Cord$, we can recognize $n \geq \alpha$ -- for this, there have to be $n - 1$ nested occurrences of the successor operation in the name of $\alpha$, and these are contained in some finite prefix. Having $n$ given as the limit of a decreasing sequence instead does not cause problems, as any premature acceptances stay valid.

Next, we shall argue that the inverse is computable, too. This means to argue that $\min : \mathcal{O}(\Cord) \to \overline{\mathbb{N}}_>$ is computable. Given an open set $U \in \mathcal{O}(\Cord)$, we can test for all natural numbers simultaneously whether $n \in U$. Any positive answer gives an upper bound for the minimal number in $U$.

Finally we need to show that the map is surjective. As $\sup : \Cord^\mathbb{N} \to \Cord$ is computable, we see that any open set has to be upwards closed. So the only thing left to argue is that any non-empty open set contains a finite ordinal, i.e.~that the finite ordinals are dense. Let us assume that an open set $U$ accepts some ordinal $\alpha$ after having read a finite prefix of its name. If every hitherto unencountered subterm in the name of $\alpha$ is replaced by $0$, the result is some finite ordinal also accepted by $U$.
\end{proof}
\end{proposition}

\begin{proposition}
The map $n \mapsto \{\alpha \in \cord \mid \alpha \geq n\} : \overline{\mathbb{N}}_< \to \mathcal{K}(\Cord)$ is a computable isomorphism.
\begin{proof}
This follows from basic properties of $\overline{\mathbb{N}}_<$.
\end{proof}
\end{proposition}

Note that in particular, the open and compact subsets of $\Cord$ coincide extensionally, thus we find $\Cord$ to be Noetherian. While this coincidence is not computable, this is only happens in trivial cases anyway. In fact, $\Cord$ is a $\nabla$-computably Noetherian space in the sense of \cite{paulydebrecht3}.

Finally, in preparation for later proofs we will explore the connection between $\mathbb{N}_<$, $\Cord$ and $\overline{\mathbb{N}}_<$ a bit more:

\begin{proposition}
\label{prop:nbar}
The following maps are computable:
\begin{enumerate}
\item $\pi : \Cord \to \overline{\mathbb{N}}_<$ acting like $\id$ on the finite ordinals, and mapping all infinite ordinals to $\infty$.
\item $\min : \overline{\mathbb{N}}_< \times \Cord \to \Cord$.
\end{enumerate}
\begin{proof}
To compute $\pi$, we simply need to count the nesting depths of $\mathalpha{+1}$ occurrences in the input ordinal.

To compute $\min : \overline{\mathbb{N}}_< \times \Cord \to \Cord$, we mostly copy the ordinal input, but under the constraint that we never allow a nesting depth of $\mathalpha{+1}$ beyond the value provided as the first component of the input. As we can assume all occurrences of $\mathalpha{+1}$ to be within the scope of a $\sup$, we can simply delay any such occurrence by inserting a number of $0$ arguments instead.
\end{proof}
\end{proposition}

\subsection{Further computable operations on $\Cord$}
We start with some basic ordinal arithmetic. Unlike the setting of reverse mathematics, we have the full power of classical logic available to prove correctness of the constructions, just the constructions themselves need to be effective. Thus, the following proposition essentially already follows from the considerations of ordinal arithmetic in reverse mathematics by \name{Hirst} \cite{hirst,hirst3,hirst2}.

\begin{proposition}
\label{prop:operations}
The following operations are computable:
\begin{enumerate}
\item $\mathalpha{+} : \Cord \times \Cord \to \Cord$
\item $\mathalpha{\times}: \Cord \times \Cord \to \Cord$
\item $(\mathalpha{-1}) : \Cord \to \Cord$, where $(\mathalpha{-1})\left (\alpha + 1\right ) = \alpha$ and for limit ordinals $\gamma$, $(\mathalpha{-1})(\gamma) = \gamma$
\item $(\alpha, \beta) \mapsto \alpha^\beta : \Cord \times \Cord \to \Cord$
\end{enumerate}
\begin{proof}
\begin{enumerate}
\item By induction on the second argument: $\alpha + 0 = \alpha$, $\alpha + \left (\beta + 1\right ) = \left ( \alpha + \beta \right ) + 1$, $\alpha + \left (\sup_{i \in \mathbb{N}} \beta_i \right ) = \sup_{i \in \mathbb{N}} \left (\alpha + \beta_i \right )$.
\item By induction on the second argument and $(1)$: $\alpha \times 0 = 0$, $\alpha \times (\beta + 1) = (\alpha \times \beta) + \alpha$, $\alpha \times (\sup_{i \in \mathbb{N}} \beta_i) = \sup_{i \in \mathbb{N}} (\alpha \times \beta_i)$.
\item Once more by induction: $(\mathalpha{-1})(0) = 0$, $(\mathalpha{-1})(\alpha + 1) = \alpha$, $(\mathalpha{-1})\left (\sup_{i \in \mathbb{N}} \alpha_i \right ) = \sup_{i \in \mathbb{N}} \left ((\mathalpha{-1})(\alpha_i) \right )$.
\item Induction on the second argument: $\alpha^0 = 1$, $\alpha^{\beta + 1} = \alpha^\beta \times \alpha$ (using 2)), $\alpha^{\sup_{i \in \mathbb{N}} \beta_i} = \sup_{i \in \mathbb{N}} \alpha^{\beta_i}$.
\end{enumerate}
\end{proof}
\end{proposition}

We can define substraction of ordinals by letting $\alpha - \beta$ be the least ordinal $\gamma$ such that $\gamma + \beta \geq \alpha$. This, however, does not yield a computable operation:

\begin{proposition}
$\lpo^* \leqW \left ( \mathalpha{-} : \Cord \times \Cord \to \Cord \right )$
\begin{proof}
For any fixed $N \in \mathbb{N}$ we could use $\mathalpha{-}$ to compute $\id : \{0,\ldots,N\}_< \to \{0,\ldots,N\}$ using the inclusion from Observation \ref{obs:finiteordinals}.
\end{proof}
\end{proposition}

Our next goal will be to understand the behaviour of suprema of continuous functions from Baire space (or some other complete space) into the countable ordinals. It is folklore that every such function is bounded by a countable ordinal. Our results provide the uniform counterpart, but we also show that $\sup : \mathcal{C}(\Baire,\Cord) \to \Cord$ is not computable.

\begin{definition}
\label{def:ev}
An \emph{evaluation tree} is a special kind of well-founded labeled tree, where each vertex has type $i$ (increment), $p$ (partition) or $s$ (supremum), and $i$ and $s$ vertices are additionally labeled with a finite word over $\mathbb{N}$. Vertices of type $i$ have exactly one child, $s$ and $p$ vertices countably many. Vertices of type $p$ on the one hand, and $i$ and $s$ on the other alternate. For any given $p$ vertex, the labels of its children never share a prefix. Let $\mathrm{ET}$ be the set of evaluation trees.
\end{definition}

\begin{definition}
Let $T$ be an evaluation tree and $p \in \Baire$. The induced ordinal $T(p)$ is defined by induction over $T$ as follows:
\begin{enumerate}
\item If the root of $T$ is an $i$-vertex, and the subtree beneath it is $T'$, then $T(p) = T'(p) + 1$.
\item If the root of $T$ is an $s$-vertex, with subtrees $(T_i)_{i \in I}$ beneath it, then $T(p) = \sup_{i \in I} T_i(p)$.
\item If the root of $T$ is a $p$-vertex, and no child of the root is labeled with a prefix of $p$, then $T(p) = 0$.
\item If the root of $T$ is a $p$-vertex, $p = wq$, and the root has a child labeled $w$ with corresponding subtree $T'$, then $T(p) = T'(q)$.
\end{enumerate}
\end{definition}

The following lemma shows that we can think of continuous functions from $\Baire$ to $\Cord$ as being represented via some evaluation tree.

\begin{lemma}
\label{lemma:evaluationtree}
The map $T \mapsto (p \mapsto T(p)) : \mathrm{ET} \to \mathcal{C}(\Baire, \Cord)$ is a computable surjection with a computable multivalued inverse.
\begin{proof}
Computability of $T(p)$ from $T$ and $p$ follows directly from the definition. We proceed to describe how to compute a total multivalued inverse. Given some realizer $F$ of continuous $f : \Baire \to (\cord,\delta_{\textrm{nK}})$, we will construct an evaluation tree $T_F$. The root will be a $p$-vertex. By observing the execution of $F$, we obtain a (prefix-independent) list of all the finite prefixes of the input causing the first digit of the input to be written.

We ignore those where this digit is $0$. For every $w$ causing the first digit to be $1$, we add an $i$-vertex with label $w$, and consider the continuous function $F' : \Baire \to \Baire$ with $F(wp) = 1F'(p)$. We iteratively apply the construction to $F'$ to obtain the subtree rooted below the $i$-vertex.  For every $w$ causing the first digit to be $2$, we add an $s$-vertex with label $w$, and consider the continuous functions $F_i$ with $F(wp) = 2\langle F_1(p),F_2(p),\ldots \rangle$. We iteratively apply the construction to each $F_i$ to obtain the subtrees rooted at the children of the $s$-vertex.

To show that this indeed defines an evaluation tree, we need to argue that the resulting tree is well-founded. Assume that the resulting tree would have some infinite path. By construction, every other vertex on that infinite path is a $p$-vertex, hence carries some label. The concatenation of all labels along the infinite path yields some $p \in \Baire$, such that $F(p)$ is not a valid $\delta_{\textrm{nK}}$-name, hence there cannot be such an infinite path if $F$ was indeed the realizer of some $f : \Baire \to (\cord,\delta_{\textrm{nK}})$.

To see that $\delta_\textrm{nK}(F(p)) = T_F(p)$ is straight-forward.
\end{proof}
\end{lemma}

\begin{definition}
An evaluation tree $T$ is said to \emph{hereditarily attain its maximum}, if for any subtree $T'$ (including $T$ itself) the function $p \mapsto T'(p) : \Baire \to \Cord$ attains its maximum.
\end{definition}

\begin{definition}
Given an evaluation tree $T$, let its bounding ordinal $\beth(T)$ be defined inductively via $\beth(T) = \sup_{i \in I} \beth(T_i)$, if the root of $T$ is a $p$-vertex or an $s$-vertex, and the subtrees rooted at its children are $(T_i)_{i \in \mathbb{N}}$, and $\beth(T) = \beth(T') + 1$, if the root of $T$ is an $i$-vertex, and $T'$ is the subtree rooted at the unique child of the root.
\end{definition}

\begin{theorem}
\label{theo:newsup}
$\beth : \mathrm{ET} \to \Cord$ is computable, and satisfies $\beth(T) \geq \sup_{p \in \Baire} T(p)$. Moreover, if $T$ hereditarily attains its maximum, then $\beth(T) = \sup_{p \in \Baire} T(p)$.
\begin{proof}
That $\beth$ is computable follows from its inductive definition. If $f(p) = \sup_{i \in \mathbb{N}} f_i(p)$, then $\sup f = \sup_{i \in \mathbb{N}} \sup f_i$. If $f(w_ip) = f_i(p)$ for some pairwise prefix-independent $(w_i)_{i \in \mathbb{N}}$, then $\sup f = \sup_{i \in \mathbb{N}} \sup f_i$. If $f(p) = f'(p) + 1$, then $\sup f \leq (\sup f') + 1$, with equality holding iff $f'$ attains its maximum. This establishes the remainder of the claim.
\end{proof}
\end{theorem}

\begin{corollary}
\label{corr:newsup}
The map $\operatorname{Bound} : \mathcal{C}(\Baire,\Cord) \mto \Cord$ where $\beta \in \operatorname{Bound}(f)$ iff $\forall p \in \Baire \ f(p) \leq \beta$ is computable.
\end{corollary}

\begin{corollary}
Let $f : \Baire \to \Cord$ be computable. Then $(\sup f) \in \Cord$ is well-defined and computable.
\begin{proof}
Combine Corollary \ref{corr:newsup} and the computability of the map $\textrm{Lower}$ established in Proposition \ref{prop:lower}.
\end{proof}
\end{corollary}

\begin{proposition}
$\sup : \mathcal{C}(\Baire,\Cord) \to \Cord$ is not computable.
\begin{proof}
Given some tree $T \subseteq \mathbb{N}^*$, let $f_T : \Baire \to \Cord$ be defined as follows: If $p \in [T]$, then $f(p) = \omega$. If $p \notin [T]$, then $f(p) = \min \{n \in \mathbb{N} \mid p_{\leq n} \notin T\}$. It is easy to see that $f_T$ is a continuous function, and is computable as such from $T$. :Let $f'_T : \Baire \to \Cord$ be defined via $f'_T(p) = f(p) + 1$. Clearly, also $f'_T$ is continuous and computable from $T$.

Now we find that $\sup f'_T = \omega + 1$, if $T$ is ill-founded, and $\sup f'_T \leq \omega$ else. In Proposition \ref{prop:leq} we will see that whether an ordinal is equal to $\omega + 1$ or less-or-equal than $\omega$ is distinguishable as a Borel truth value. However, whether a tree is ill-founded or not is the canonical example of a $\Sigma^1_1$-complete problem, hence distinguishing the two cases cannot be done in a Borel way. The only way to resolve the apparent contradiction is to realize that $\sup : \mathcal{C}(\Baire,\Cord) \to \Cord$ is not computable.
\end{proof}
\end{proposition}

In order to prove some more involved computability results, a normal form of $\delta_{\textrm{nK}}$-names will be very useful. Intuitively, a normal form name does not contain immediate nestings of $\sup$.

\begin{definition}
\label{def:naivekleenenf}
We define a representation $\delta_{\textrm{nK}}^{\textrm{nf}} : \subseteq \Baire \to \cord$ inductively via:
\begin{enumerate}
\item $\delta_{\textrm{nK}}^{\textrm{nf}}(0p) = 0$
\item $\delta_{\textrm{nK}}^{\textrm{nf}}(1p) = \delta_{\textrm{nK}}^{\textrm{nf}}(p) + 1$
\item $\delta_{\textrm{nK}}^{\textrm{nf}}(2\langle p_0, p_1, p_2, \ldots \rangle) = \sup_{i \in \mathbb{N}} \delta_{\textrm{nK}}^{\textrm{nf}}(p_i)$, provided that $\forall i \in \mathbb{N} \ p_i(0) \in \{0,1\}$.
\end{enumerate}
\end{definition}

\begin{observation}
$\Cord \cong (\cord, \delta_{\textrm{nK}}^{\textrm{nf}})$.
\begin{proof}
The identity $\id : (\cord, \delta_{\textrm{nK}}^{\textrm{nf}}) \to \Cord$ is realised by the identity on $\Baire$. For the other direction we simply use tupling functions to flatten out immediate nestings of $\sup$.
\end{proof}
\end{observation}

\begin{observation}
\label{obs:domdeltanf}
$\dom(\delta_{\textrm{nK}}^{\textrm{nf}})$ is a closed subset of $\dom(\delta_{\textrm{nK}})$, and not even a $\bolds^1_1$ subset of $\Baire$.
\begin{proof}
If $p \in \dom(\delta_{\textrm{nK}}) \setminus \dom(\delta_{\textrm{nK}}^{\textrm{nf}})$, then there is some immediate nesting of $\sup$ in $p$, which we can detect. For the second claim, note that a Wadge-reduction from the set of well-founded trees can be obtained in a straight-forward manner, and this is the standard example of a $\Pi^1_1$-complete set, see e.g.~\cite{moschovakisb}.
\end{proof}
\end{observation}

\begin{definition}
\label{def:leqcertificate}
Given some $q \in \dom(\delta_{\textrm{nK}})$ and some $p \in \Baire$, we call $r \in \Baire$ a $\leq$-certificate for $(p,q)$ iff the following conditions are satisfied:
\begin{enumerate}
\item If $q(0) = 0$, then $p(0) = 0$.
\item If $q = 1q_0$, then either $p = 1p_0$ and $r$ is a $\leq$-certificate for $(p_0,q_0)$, or $p = 2\langle n_0p_0,n_1p_1,\ldots\rangle$ with $n_i \in \{0,1\}$ and $r = \langle r_0,\ldots\rangle$, and for each $i \in \mathbb{N}$ with $n_i = 1$, $r_i$ is a $\leq$-certificate for $(p_i,q_0)$.
\item If $q = 2\langle q_0,q_1,\ldots\rangle$, then either
\begin{enumerate}
\item $p(0) = 0$.
\item $p = 1p_0$, $r = nr_0$ and $r_0$ is a $\leq$-certificate for $(p, q_n)$.
\item or $p = 2\langle p_0, p_1,\ldots\rangle$, $r = \langle h, \langle r_0,r_1,\ldots\rangle\rangle$, and for each $i \in \mathbb{N}$, $r_i$ is a $\leq$-certificate for $(p_i,q_{h(i)})$.
\end{enumerate}
\end{enumerate}
\end{definition}

\begin{observation}
\label{obs:leqcertificate}
Let $q \in \dom(\delta_{\textrm{nK}})$. If $(p,q)$ has a $\leq$-certificate, then $p \in \dom(\delta_{\textrm{nK}})$ and $\delta_{\textrm{nK}}(p) \leq \delta_{\textrm{nK}}(q)$. If $p \in \dom(\delta_{\textrm{nK}}^{\textrm{nf}})$ and $\delta_{\textrm{nK}}^{\textrm{nf}}(p) \leq \delta_{\textrm{nK}}(q)$, then $(p,q)$ admits a $\leq$-certificate. The set of $\leq$-certificates for $(p,q)$ is a closed set computable from $p$ and $q$.
\begin{proof}
By induction over the structure of $q$.
\end{proof}
\end{observation}

Based on the computability of the supremum of an evaluation tree that hereditarily attains is maximum and the notion of a $\leq$-certificate, we can prove two results about the computability of minima. The result on countable minima is weaker, but easier to obtain. It will be followed by a stronger result on binary minima.

\begin{proposition}
\label{prop:countablemin}
The map $(\alpha_i)_{i \in \mathbb{N}} \mapsto \max \{\omega, \min_{n \in \mathbb{N}} \alpha_n\} : \mathcal{C}(\mathbb{N},\Cord) \to \Cord$ is computable.
\begin{proof}
We begin by computing a continuous function $f : \Baire \to \Cord$ from the $(\alpha_i)_{i \in \mathbb{N}}$. Let us assume we are given $\delta_{\mathrm{nK}}$-names $(q_i)_{i \in \mathbb{N}}$ for the $(\alpha_i)_{i \in \mathbb{N}}$. A realizer of $F$ proceeds as follows on input $\langle p, \langle r_0,r_1,\ldots\rangle\rangle$: Test whether for each $i \in \mathbb{N}$, $r_i$ is a $\leq$-certificate for $(p,q_i)$. As long as this is consistent, output $p$. If a counterexample is found (Observation \ref{obs:leqcertificate}), complete the partial output into a name of some finite ordinal.

We claim that $\min_{n \in \mathbb{N}} \alpha_n \leq \sup_{p \in \Baire} f \leq  \max \{\omega, \min_{n \in \mathbb{N}} \alpha_n\}$. Let $p$ be some $\delta_{\mathrm{nK}}^{\mathrm{nf}}$-name of some ordinal $\beta \leq \min_{n \in \mathbb{N}}$. Then by Observation \ref{obs:leqcertificate}, for each $i \in \mathbb{N}$ there is some $\leq$-certificate $r_i$ for $(p, q_i)$, and thus, $\beta$ will occur in the range of $f$. Conversely, if some infinite $\beta$ occurs in the range of $f$, corresponding $\leq$-certificates $(r_i)$ must exist, and hence $\beta \leq \min_{n \in \mathbb{N}} \alpha_n$.

Finally, we argue that the evaluation tree obtained by Lemma \ref{lemma:evaluationtree} from the realizer constructed for $f$ hereditarily attains its maximum. As shown above, $f$ itself attains its maximum. By aligning Definition \ref{def:leqcertificate} with Lemma \ref{lemma:evaluationtree} we see that the function computed by any subtree of the evaluation tree of $f$ either corresponds to the function that we would construct from another sequence $(\alpha'_i)$ -- hence also attains its maximum, or is constant with a finite value. We can thus use Theorem \ref{theo:newsup} to compute $\sup f$, and then $\max \{\sup f, \omega\} = \max \{\omega, \min_{n \in \mathbb{N}} \alpha_n\}$.
\end{proof}
\end{proposition}

\begin{theorem}
\label{theo:binarymin}
$\min : \Cord \times \Cord \to \Cord$ is computable.
\begin{proof}
Combine Proposition \ref{prop:nbar} and Proposition \ref{prop:countablemin}: Given inputs $\alpha, \beta \in \Cord$, we can compute $\gamma := \max \{\omega, \min \{\alpha, \beta\}\} \in \Cord$ and $\pi(\alpha), \pi(\beta) \in \overline{\mathbb{N}}_<$. Then we compute $\min \{ \pi(\alpha), \min \{\pi(\beta), \gamma\}\} \in \Cord$ and find this to be equal to $\min \{\alpha, \beta\}$.
\end{proof}
\end{theorem}

Note that the proof of Proposition \ref{prop:countablemin} and Theorem \ref{theo:binarymin} actually gives us more: Let $\overline{\Cord}$ be the extension of $\Cord$ by $\infty$ (which is assumed to be greater than any ordinal), where we let $p$ represent $\infty$ iff $p$ contains an infinite nesting of $1$s. Then we obtain:

\begin{proposition}
\label{prop:minext}
$\min : \subseteq \mathcal{C}(\mathbb{N},\overline{\Cord}) \to \Cord$ with $\dom(\min) = \{(\alpha_i)_{i \in \mathbb{N}} \mid \exists i \ \alpha_i < \infty \wedge \forall j \ \alpha_j \geq \omega\}$ is computable.
\end{proposition}

\begin{proposition}
\label{prop:binaryminext}
$\min : \subseteq \overline{\Cord} \times \overline{\Cord} \to \Cord$ with $\dom(\min) = \{(\alpha, \beta) \in \overline{\Cord}^2 \mid \alpha < \infty \vee \beta < \infty\}$ is computable.
\end{proposition}

\subsection{From Borel to continuous}
\label{subsec:borel}
We will show that, in some sense, Borel maps into $\Cord$ are not much more powerful than continuous maps. Our proof will use the following lemma on extending partial functions $f : \subseteq \Baire \to \Cord$:

\begin{lemma}
\label{lem:extension}
Given $A \in \mathcal{A}(\Baire)$ and $f \in \mathcal{C}(A,\Cord)$ we can compute some $g \in \mathcal{C}(\Baire, \Cord)$ such that $g|_A = f$ and for $p \in \Baire \setminus A$, $g(p)$ is a finite ordinal.
\begin{proof}
Being given $f \in \mathcal{C}(A,\Cord)$ means we have some realizer $F$ of it available. We can attempt to apply $F$ to some $p \notin A$, and can w.l.o.g. assume that $F$ will only output digits $0$, $1$ or $2$ -- thus any finite prefix of the output is the prefix of some name for a (finite) element of $\Cord$. Of course $F$ may fail to produce any output. Being given $A \in \mathcal{A}(\Baire)$ means we can semidecide whether $p \notin A$. Thus, we obtain a realizer $G$ for $g$ by running $F$ while testing $p \notin A$. If the latter is ever confirmed, we complete the output of $F$ so far into a name for some finite ordinal. Otherwise, $F$ will produce the correct output.
\end{proof}
\end{lemma}

To formulate the main result of this section, we need the closed choice principle for Baire space $\C_\Baire : \subseteq \mathcal{A}(\Baire) \mto \Baire$. Introduced in \cite{brattka3}, $\C_\Baire$ takes a non-empty closed subset $A$ of $\Baire$ as input, and outputs some element $p \in A$.

\begin{theorem*}[\name{Brattka}, \name{de Brecht} \& \name{P.} \cite{paulybrattka}]
Let $\mathbf{X}$, $\mathbf{Y}$ be computable Polish spaces. Then the following are equivalent for $f : \mathbf{X} \to \mathbf{Y}$:
\begin{enumerate}
\item $f$ is Borel measurable.
\item $f \leqW \C_\Baire$ relative to some oracle.
\end{enumerate}
\end{theorem*}

Based on the preceding result, it makes sense to consider Weihrauch reducibility to $\C_\Baire$ as a generalization of Borel measurability in settings where the latter is not a agreed upon notion -- such as $\Cord$. This could be formalized by deriving a computable endofunctor from $\C_\Baire$ along the lines of \cite{pauly-descriptive-lics,paulydebrecht2}.

\begin{theorem}
\label{theo:cbaireremoval}
Let $f : \Baire \to \Cord$ satisfy $f \leqW \C_\Baire$. Then there is a computable function $g : \Baire \to \Cord$ with $\forall p \in \Baire \ \ f(p) \leq g(p)$.
\begin{proof}
The reduction $f \leqW \C_\Baire$ means that given some $\delta_\mathbf{X}$-name $p$, we can compute some non-empty $A_p \in \mathcal{A}(\Baire)$ and a continuous function $K_p : A_p \to \Cord$ such that $K_p(q) = f(\delta_\mathbf{X}(p))$ for any $q \in A_p$. We then use Lemma \ref{lem:extension} to obtain some extension $G_p : \Baire \to \Cord$ of $K_p$. After that, we can invoke Corollay \ref{corr:newsup} to obtain some upper bound $g(p)$ for $\sup K_p$.
\end{proof}
\end{theorem}

\begin{corollary}
\label{corr:cbaireremoval}
Let $f : \Baire \to \Cord$ satisfy $f \leqW \C_\Baire$. Then $\sup_{p \in \Baire} f(p)$ is a computable countable ordinal.
\end{corollary}

\begin{corollary}
Let $f : \Baire \to \Cord$ admit a(n effectively) Borel-measurable realizer. Then $\sup_{x \in \mathbf{X}} f(x)$ is a (computable) countable ordinal.
\begin{proof}
By the result from \cite{paulybrattka}, we can apply (the relativization of) Theorem \ref{corr:cbaireremoval} to the realizer of $f$, and then Corollay \ref{corr:newsup} to the result.
\end{proof}
\end{corollary}

\section{Ordinals in other structures}
\label{sec:otherstructures}
In this section we will consider several ordinal-properties of various structures, and investigate whether they are computable maps from the corresponding spaces into $\Cord$, and whether these maps admit computable multivalued inverses. Any of these results can also be read as introducing another representation of the countable ordinals (via coding an ordinal $\alpha$ as a suitable structure with $\alpha$ as the value of the property), together with a proof or disproof of equivalence with the standard representation.

\subsection{Height of posets}
\label{subsec:posets}
Let us consider partially order sets $(A,\prec)$ with $A \subseteq \mathbb{N}$. We can represent these by coding the characteristic function of $A$ and and of $\mathalpha{\prec} \subseteq \mathbb{N} \times \mathbb{N}$ into some $p \in \Cantor$. We are only interested in well-founded relations, i.e.~those without infinite decreasing sequences. Let $\mathcal{R}_{\textrm{wf}}$ denote the represented space of well-founded relations. We call $(A, \prec)$ a (set-theoretic) tree, if for any $n \in A$ we find that $(\{t \in A \mid t \prec n\}, \prec)$ is a chain. Let $\mathcal{R}_{\textrm{tr}}$ denote the represented space of set-theoretic trees.

Recall that a chain is some $B \subseteq A$ such that for $x \neq y \in B$ either $x \prec y$ or $y \prec x$. The order type of a chain is some countable ordinal. The height of $(A, \prec)$ is the supremum of the order types of all chains in $(A, \prec)$.

\begin{theorem}
\label{theo:height}
\begin{enumerate}
\item $\operatorname{height} : \mathcal{R}_{\textrm{wf}} \to \Cord$ is computable.
\item $\operatorname{height}^{-1} : \Cord \mto \mathcal{R}_{\textrm{wf}}$ is computable.
\item $\operatorname{height} : \mathcal{R}_{\textrm{tr}} \to \Cord$ is computable.
\item $\operatorname{height}^{-1} : \Cord \mto \mathcal{R}_{\textrm{tr}}$ is computable.
\end{enumerate}
\begin{proof}
\begin{enumerate}
\item For $n \in A$, let $A_n := \{t \in A \mid t \prec n\}$. Any chain in $A_n$ can be extended by $n$ to yield a chain in $A$, thus we find that $\operatorname{height}(A_n) + 1 \leq \operatorname{height}(A)$. If $B \subseteq A$ is a chain, then there is some co-final sequence $(b_i)_{i \in \mathbb{N}}$ in $B$, which means that the order type of $B$ is the supremum of the order types of $\{t \in B \mid t \preceq b_i\}$. We thus find that $\operatorname{height}(A) = \sup_{n \in A} (\operatorname{height}(A_n) + 1)$. By setting $\alpha_n := \operatorname{height}(A_n) + 1$ if $n \in A$ and $\alpha_n := 0$ else, we can compute $\operatorname{height}(A)$ as $\sup_{n \in \mathbb{N}} \alpha_n$ by induction.

\item As $\mathcal{R}_{\textrm{tr}}$ is a subspace of $\mathcal{R}_{\textrm{wf}}$, and $\operatorname{height} : \mathcal{R}_{\textrm{tr}} \to \Cord$ is already surjective, this follows from $(4.)$.

 \item Trivial consequence of $(1.)$.

\item We use the argument of Proposition \ref{prop:nkinitial}. The empty relation is a computable element of $\mathcal{R}_{\textrm{tr}}$ and has height $0$.

For $\mathalpha{+1}$, let us be given a tree $(A,\prec) \in \mathcal{R}_{\textrm{tr}}$. Let $A' = \{2n \mid n \in A\} \cup \{2n + 1 \mid n \in A\}$. Let $\prec'$ be defined via $2n \prec' 2k$ iff $n \prec k$, $2n \prec' 2k+1$ iff $n \prec k$ or $n = k$ and $2k+1 \nprec' x$. Intuitively, we add a new element above each element already present. This yields a tree again, and $\operatorname{height}(A',\prec') = \operatorname{height}(A,\prec) + 1$.

For $\sup$, note that we can use a tupling function on $\mathbb{N}$ to make sense of countable disjoint unions, and that $\mathcal{R}_{\textrm{tr}}$ is then closed under countable disjoint unions in an effective way. The height of the disjoint union of the posets $A_k$ is just $\sup_{k \in \mathbb{N}} \operatorname{height}(A_k)$. Invoking Proposition \ref{prop:nkinitial} now yields computability of $\operatorname{height}^{-1}$.
\end{enumerate}
\end{proof}
\end{theorem}

Note that the preceding theorem is unaffected if we replace the characteristic functions of $A$ and $\prec$ by enumerations of the corresponding sets. If, however, we would use enumerations of their complements instead, the situation differs on the finite ordinals. For $X, Y \in \{\mathcal{A},\mathcal{O},\mathcal{O}\wedge\mathcal{A}\}$ let $\mathcal{R}_{\textrm{wf}}^{X,Y}$ denote the set of well-founded posets $(A, \prec)$ where $A$ is given as an enumeration if $X = \mathcal{O}$, as an enumeration of its complement if $X = \mathcal{A}$ and as a characteristic function if $X = \mathcal{O} \wedge \mathcal{A}$. Likewise the value of $Y$ denotes how $\prec$ is given. In particular, $\mathcal{R}_{\textrm{wf}} := \mathcal{R}_{\textrm{wf}}^{\mathcal{O}\wedge\mathcal{A},\mathcal{O}\wedge \mathcal{A}}$. Now $\mathcal{R}_{\textrm{tr}}^{X,Y}$ is just the restriction of $\mathcal{R}_{\textrm{wf}}^{X,Y}$ to set-theoretic trees. For any of these spaces $\mathcal{R}$, let $\mathcal{R}|_\mathcal{F}$ denote its restriction to structures of finite height.

\begin{corollary}
The following maps are computable for $X,Y \in \{\mathcal{O},\mathcal{O}\wedge\mathcal{A}\}$:
\begin{enumerate}
\item $\operatorname{height} : \mathcal{R}_{\textrm{wf}}^{X,Y} \to \Cord$.
\item $\operatorname{height}^{-1} : \Cord \mto \mathcal{R}_{\textrm{wf}}^{X,Y}$.
\item $\operatorname{height} : \mathcal{R}_{\textrm{tr}}^{X,Y} \to \Cord$.
\item $\operatorname{height}^{-1} : \Cord \mto \mathcal{R}_{\textrm{tr}}^{X,Y}$.
\end{enumerate}
\begin{proof}
By the proof of Theorem \ref{theo:height}.
\end{proof}
\end{corollary}

In order to formulate the following proposition, we will need the space $\mathbb{N}_<'$. Here names are sequences converging to a name for the same element of $\mathbb{N}_<$. An alternative equivalent, maybe more intuitive, representation is that $p \in \Baire$ codes some element in $\mathbb{N}_<'$ iff not all numbers occur infinitely often in the range of $p$, and then it codes the largest natural number occurring only finitely many times.

\begin{proposition}
Let $X,Y \in \{\mathcal{O},\mathcal{A},\mathcal{O} \wedge \mathcal{A}\}$. Then the following maps are computable:
\begin{enumerate}
\item $\operatorname{height} : \mathcal{R}_{\textrm{wf}}^{X,Y}|_\mathcal{F} \to \mathbb{N}_<'$.
\item $\operatorname{height}^{-1} : \mathbb{N}_<' \mto \mathcal{R}_{\textrm{wf}}^{\mathcal{A},Y}$.
\item $\operatorname{height} : \mathcal{R}_{\textrm{tr}}^{X,Y}|_\mathcal{F} \to \mathbb{N}_<'$.
\item $\operatorname{height}^{-1} : \mathbb{N}_<' \mto \mathcal{R}_{\textrm{tr}}^{\mathcal{A},Y}$.
\item $\max \{1, \operatorname{height}\}^{-1} : \mathbb{N}_<' \mto \mathcal{R}_{\textrm{tr}}^{X,\mathcal{A}}$.
\item $\max \{1, \operatorname{height}\}^{-1} : \mathbb{N}_<' \mto \mathcal{R}_{\textrm{wf}}^{X,\mathcal{A}}$.
\end{enumerate}
\begin{proof}
\begin{enumerate}
\item Let the finite posets $A_t$ be the approximations to the input made available at stage $t$. For each $n \in \mathbb{N}$ we keep track of whether there is a chain of length $n$ present in $A_t$. If there is none, or if our current specific candidate of such a chain was not already present in $A_{t-1}$, we write $n$ as part of the output. We chose the candidate chains in some linearly ordered way.

    Consider some maximal chain $c$ in $A$. Then $c$ will be present in all but finitely many $A_t$. Either it or some other maximal chain in $A$ of the same length will eventually become the current candidate, and thereafter we will no longer write $|c|$. Conversely, if there is no chain of length $n$ in $A$, then no chain of length $n$ can occur in infinitely many $A_t$. Thus, we will write $n$ infinitely many times.
\item This will follow from $(4)$.
\item This follows from $(1)$.
\item For each $n \in \mathbb{N}$, we create a separate chain of length $n$. Whenever we read $n$ in the $\mathbb{N}_<'$-name we receive as input, we remove the support of the last chain of length $n$ from $A$, and create a new such chain. The final output will have a chain of length $n$ iff $n$ occurs only finitely many times in the input, which already ensures correctness.
\item Similar to the construction in $(4)$, except that rather than removing the chains entirely, we turn them into antichains by removing the suitable pairs from $\prec$. These relics then contribute a height of at least $1$, but are unproblematic beyond that.
\item This is a consequence of $(5)$.
\end{enumerate}
\end{proof}
\end{proposition}

\subsection{Heights of wellorders}
Often, ordinals are consider to be the order-types of wellorders. Similar to the approach explored in Subsection \ref{subsec:posets}, we can then introduce computability on $\cord$ by restricting $\mathcal{R}_{\textrm{wf}}$ further to wellorders $\mathcal{R}_{\textrm{wo}}$, and then identifying isomorphic orders. Computability on the resulting space was studied by Joel \name{Hamkins} and Zhenhao \name{Li} in \cite{zhenhao}, and we shall thus name it $\CordHL$. We briefly survey some of their results:

\begin{theorem}[Hamkins \& Li \cite{zhenhao}]
The following operations are computable:
\begin{enumerate}
\item $\mathalpha{+} : \CordHL \times \CordHL \to \CordHL$
\item $\mathalpha{\times} : \CordHL \times \CordHL \to \CordHL$
\item $(\alpha, \beta) \mapsto \alpha^\beta : \CordHL \times \CordHL \to \CordHL$
\item $\alpha + 1 \mapsto \alpha :\subseteq \CordHL \to \CordHL$
\item $\omega^{\textrm{CK}} + \omega \mapsto \omega^{\textrm{CK}} : \subseteq \CordHL \to \CordHL$
\end{enumerate}
\end{theorem}

The first item of the following justifies our rejection of $\CordHL$ as proposed \emph{standard computability structure} on the countable ordinals. We point out that the technique introduced in \cite[Theorem 16]{zhenhao} essentially is a Wadge game relative to the representation, similar to the generalizations of the classical Wadge hierarchy on $\Baire$ to represented spaces in \cite{pequignot} by \name{Pequignot} and \cite{duparc} by \name{Duparc} and \name{Fournier}.

\begin{theorem}[Hamkins \& Li \cite{zhenhao}]
The following operations are not computable:
\begin{enumerate}
\item $\max :\CordHL \times \CordHL \to \CordHL$
\item $\alpha \mapsto \max \{\alpha, \omega + 1\} : \CordHL \to \CordHL$
\item $\omega \times \alpha \mapsto \alpha : \subseteq \CordHL \to \CordHL$
\item $\textrm{Reduce}_n : \subseteq \CordHL \to \CordHL$ where $\textrm{Reduce}_n(\omega) = n$ and $\textrm{Reduce}_n(\omega + \omega) = \omega$
\item $\textrm{D} : \subseteq \CordHL \to \{0,1\}$ where $\textrm{D}(\omega) = 0$ and $\textrm{D}(\omega + 1) = 1$
\end{enumerate}
\end{theorem}

Finally, we point out that the investigations in \cite[Section 5]{zhenhao} concern the \emph{point degree spectrum} of $\CordHL$ (without using this terminology, though). Point degree spectra of represented spaces were introduced by \name{Kihara} and \me in \cite{pauly-kihara-arxiv}.

It is common knowledge that almost all approaches to define computability on the countable ordinals will yield the same notion of what a computable ordinal is. For the comparison of $\Cord$ and $\CordHL$, the following provides the explanation:

\begin{proposition}
$\operatorname{LinExt} : \mathcal{R}_{tr} \mto \mathcal{R}_{\textrm{wo}}$ mapping a well-founded set-theoretic tree to some linear extension is computable.
\begin{proof}
We can use the standard order on the natural numbers to linearly order the subtrees of any particular vertex, and then use the Kleene-Brouwer ordering on the resulting tree.
\end{proof}
\end{proposition}

\begin{corollary}
The map $\operatorname{UpperBound} : \Cord \mto \CordHL$ mapping an ordinal to some ordinal as least as large in $\CordHL$ is computable.
\end{corollary}

\begin{observation}
\label{prop:lowerHL}
$\textrm{Lower} : \CordHL \mto \mathcal{C}(\mathbb{N},\{\textrm{Skip}\} \uplus \CordHL)$ is computable.
\end{observation}

\begin{corollary}
\label{corr:cordhlpoints}
$\Cord$ and $\CordHL$ have the same computable points.
\end{corollary}

\subsection{Further structures}
\label{subsec:furtherstructures}
Based on Definition \ref{def:naivekleene}, it is straight-forward that the rank of a well-founded tree is computable in $\Cord$, and that given some ordinal $\alpha$ we can compute a tree with rank $\alpha + 1$. The same argument applies to the ranks of Borel codes as used in \cite{moschovakisb}. The following presentation follows \cite{pauly-gregoriades}:

\begin{definition}(\cite{moschovakisb} 3H)
The set of \emph{Borel codes} $\bcode \subseteq \Baire$ is defined by recursion as follows
\begin{align*}
p \in \bcode_0 &\iff p(0) = 0\\
p \in \bcode_\alpha &\iff p = 1\langle p_0, p_1, \ldots, \rangle \ \& \ (\forall n)(\exists \beta < \alpha)[p_n \in \bcode_\beta]\\
\bcode = \cup_{\alpha} \bcode_\alpha& \hspace{2mm} \textrm{for all countable ordinals $\alpha$.}
\end{align*}
\end{definition}

For all $p \in \bcode$ we denote by $|p|$ its \emph{rank}, that is the least ordinal $\alpha$ such that $p \in \bcode_\alpha$. It is not hard to verify that
\[
|1\langle p_0,p_1,\ldots\rangle| = \sup_{n \in \mathbb{N}} |p_n| + 1
\]

\begin{observation}
$| \ | : \bcode \to \Cord$ is computable, and there is a computable $I : \Cord \mto \bcode$ such that $|I(\alpha)| = \alpha + 1$.
\end{observation}

Let $\mathbf{X}$ be a Polish space, and let $\mathcal{B}(\mathbf{X})$ denote the space of Borel subsets of $\mathbf{X}$. For some subset $A \subseteq \mathbf{X}$, let $A^C$ denote its complement $\mathbf{X} \setminus A$. Given some standard representation $\delta_\mathcal{O} : \Baire \to \mathcal{O}(\mathbf{X})$ of its open sets, we will define the standard representation $\pi :\subseteq \Baire \to \mathcal{B}(\mathbf{X})$ of its Borel subsets:
\begin{align*}
\pi^\mathbf{X}(0p) =& \delta_\mathcal{O}(p)\\
\pi^\mathbf{X}(1\langle p_0, p_1, \ldots\rangle) =& \bigcup_{n} \left( \pi^{\mathbf{X}}(p_n)\right)^C.
\end{align*}

Note that a set has a $\pi$-name of rank $\alpha$ iff it is a $\bolds^0_{1+\alpha}$-set. This in turn implies that given a Borel set $A \in \mathcal{B}(\mathbf{X})$ we can compute some ordinal $\alpha \in \Cord$ such that $A \in \bolds^0_\alpha$. We cannot compute a minimal such $\alpha$ though:

\begin{proposition}
The function $M : \mathcal{B}(\Baire) \mto \Cord$ mapping any $\bolds^0_{\omega+1}$-complete set to some ordinal $\alpha \geq \omega + 1$ and the empty set to some ordinal $\beta \leq \omega$ is not computable.
\begin{proof}
Fix some $\Sigma^0_{\omega+1}$-complete set $B$. Given some closed $A \in \mathcal{A}(\Baire)$, we can compute $A \times B$. This is $\bolds^0_{\omega+1}$-complete, if $A \neq \emptyset$ and $A \times B = \empty$ if $A = \emptyset$. The set $\{\beta \in \Cord \mid \beta \geq \omega + 1\}$ is $\Sigma^0_3$, hence, if $M$ were computable, then $\{A \in \mathcal{A}(\Baire) \mid A \neq \emptyset\}$ were $\bolds^0_3$, too. However, this set is known to be $\Sigma^1_1$-complete.
\end{proof}
\end{proposition}

Another setting where the countable ordinals appear is as the Cantor-Bendixson rank of closed sets. However, as shown by \name{Kreisel} \cite{kreisel}, there is a computable closed set with Cantor-Bendixson rank $\omega^{\textrm{CK}}$. Hence, the Cantor-Bendixson rank is not even non-uniformly computable. For investigations of computability-aspects relating to the Cantor-Bendixson rank we refer to \cite{cenzer2,normann}.

\section{Alternative approaches to computability on $\cord$}
\label{sec:alternatives}
In this section, we will briefly discuss some alternative approaches to define computability on $\cord$, and our reasons to reject them.
\subsection{A Kleene-style representation}
Another alternative candidate for a representation on $\cord$ is a straightforward adaption of \name{Kleene}'s notation \cite{kleene3} of the recursive ordinals to a representation of the countable ordinals. The resulting representation $\delta_{\textrm{K}}$ is a restriction of $\delta_{\textrm{nK}}$, with the additional requirement that $\sup$'s may only be taken about strictly increasing sequence\footnote{This is similar to the relationship between the Cauchy representation and the naive Cauchy representation of a computable metric space \cite{weihrauchd}: In the latter, points are represented via fast converging sequences of basic points, in the latter by any converging sequence. Hence, the naming of $\delta_{\textrm{K}}$ and $\delta_{\textrm{nK}}$. Of course, here we are arguing that $\delta_{\textrm{nK}}$ is better behaved than $\delta_{\textrm{K}}$, whereas for computable metric spaces, the Cauchy representation, not the naive Cauchy representation is the appropriate choice.}.

\begin{definition}
We define $\delta_{\textrm{K}} : \subseteq \Baire \to \cord$ inductively via: \begin{enumerate}
\item $\delta_{\textrm{K}}(0p) = 0$
\item $\delta_{\textrm{K}}(1p) = \delta_{\textrm{K}}(p) + 1$
\item $\delta_{\textrm{K}}(2\langle p_0, p_1, p_2, \ldots \rangle) = \sup_{i \in \mathbb{N}} \delta_{\textrm{K}}(p_i)$, provided that $\forall i \in \mathbb{N} \ \delta_{\textrm{K}}(p_i) < \delta_{\textrm{K}}(p_{i+1})$.
\end{enumerate}
\end{definition}

We shall see that $(\cord, \delta_{\textrm{K}})$ lacks the nice closure properties of $\Cord$:

\begin{proposition}
\label{prop:cordkmax}
$\lpo \leqW \left ( \max : (\cord, \delta_{\textrm{K}}) \times (\cord, \delta_{\textrm{K}}) \to (\cord, \delta_{\textrm{K}}) \right )$
\begin{proof}
We start with the observation that both $\iota : \mathbb{S} \to (\cord, \delta_{\textrm{K}})$ defined via $\iota(\bot) = \omega$ and $\iota(\top) = 2\omega$ as well as the constant $\omega + 1$ are computable. Then note that being a limit ordinal is decidable on $(\cord, \delta_{\textrm{K}})$, yielding a computable function $\operatorname{IsLimit} : (\cord, \delta_{\textrm{K}}) \to \mathbf{2}$. Now $\mathbf{t} \mapsto \operatorname{IsLimit}(\max (\iota(\mathbf{t}), \omega + 1))$ is identical to $\id : \mathbb{S} \to \mathbf{2}$.
\end{proof}
\end{proposition}

Just as we saw in Corollary \ref{corr:cordhlpoints} for $\CordHL$, we can see that we can make the additional requirements for $\delta_{\textrm{K}}$-names true by moving to a larger ordinal:

\begin{proposition}
\label{prop:upperbound}
The map $\operatorname{UpperBound} : \Cord \mto (\cord,\delta_{\textrm{K}})$ defined by $\beta \in \operatorname{UpperBound}(\alpha)$ iff $\beta \geq \alpha$ is computable.
\begin{proof}
The computation proceeds by induction, using the representations $\delta_{\textrm{nK}}$ and $\delta_{\textrm{K}}$. For $0$ and successor, both representations agree anyway. Given a supremum $\alpha = \sup_{n \in \mathbb{N}} \alpha_n$, we apply $\operatorname{UpperBound}$ to each $\alpha_n$ to obtain an upper bound $\beta_n$. Now $\beta = \sup_{n \in \mathbb{N}} \left (\beta_0 + \ldots \beta_n \right )$ is a valid output for $\operatorname{UpperBound}(\alpha)$ (note that addition is computable on $(\cord,\delta_{\textrm{K}})$).
\end{proof}
\end{proposition}

\begin{corollary}
The computable elements of $\Cord$ and $(\cord,\delta_{\textrm{K}})$ are the same.
\end{corollary}

The advantage one might see in $(\cord,\delta_{\textrm{K}})$ is that there the set of limit ordinals becomes a decidable subset. Given that case distinctions between limit and successor ordinals are ubiquitous in definitions concerning ordinals, this might seem very desirable. However, these definitions can generally be adapted to work with $\delta_{\textrm{nK}}$-names instead: One just needs to ensure that the limit-case is not itself adding complexity, but merely gathering the complexity added in the preceding successor-steps. Then unnecessary applications of the limit step are harmless.

For example, we defining the Borel sets, we would define $\bolds^0_{\alpha+1}$ to be the countable unions of complements of $\bolds^0_{\alpha}$-sets and $\bolds^0_{\sup_{n} \gamma_n} := \bigcup_{n \in \mathbb{N}} \bolds^0_{\gamma_n}$. Note though that this differs by $1$ from the usual definition for infinite ordinals.

\subsection{A non-deceiving representation?}
The trusted recipe of identifying suitable representations of some structure is to pick an admissible representation whose final topology coincides with some natural topology on the structure\footnote{In fact, it is sometimes claimed that it \emph{has} to be done like that -- the present work ought to disprove this.}. However, the usual topology on $\cord$ would be the order topology, which is not separable -- and every represented space is separable. In this section, we shall explore whether a weaker topological requirement could be imposed on a representation.

Inspired by a property studied in the context of winning conditions for infinite sequential games in \cite{paulyleroux2-arxiv} by \name{Le Roux} and \name{P.}, we shall call a function $f : \subseteq \Baire \to \cord$ \emph{non-deceiving}, iff whenever $(p_n)_{n \in \mathbb{N}}$ is a sequence converging to $p$ in $\dom(f)$ such that $\forall n \in \mathbb{N} \ f(p_n) < f(p_{n+1})$, then $\forall i \in \mathbb{N} \ f(p_i) < f(p)$.

\begin{theorem}[Gregoriades\footnote{This theorem is based on a personal communication by Vassilios \name{Gregoriades}.}]
Any non-deceiving function $f : \subseteq \Baire \to \cord$ is bounded by some countable ordinal.
\begin{proof}
As $\dom(f)$ (as a subspace of $\Baire$) is countably based, it suffices to show that for any $p \in \dom(f)$ there is some open neighborhood $U$ of $p$ such that $f|_U$ is bounded (as there are only countably many basic open sets, the supremum of the local bounds is a global bound).

Assume the contrary. Then there is some $p \in \dom(f)$ such that for each $n \in \mathbb{N}$ and each $\alpha \in \cord$ there is some $q \in \dom(f)$ with $d(p,q) < 2^{-n}$ and $f(q) >\alpha$. By choosing a suitable $q$ countable many times, we arrive at a sequence $(q_i)_{i \in \mathbb{N}}$ with $\lim_{i \to \infty} q_i = p$ and $f(p) < f(q_0) < f(q_1) < \ldots$. But this contradicts the non-deceiving condition.
\end{proof}
\end{theorem}

\begin{corollary}
There is no non-deceiving representation of $\cord$.
\end{corollary}

The preceding corollary presumably destroys any hope to find a suitable represention of $\cord$ that is admissible w.r.t.~some weak limit space structure in the sense of \name{Schr\"oder} \cite{schroder4,schroder5}.

\section{Ordinals in descriptive set theory}
\label{sec:descriptive}
In this section, we shall use the space $\Cord$ in order to further develop the uniform approach to descriptive set theory on represented spaces as proposed in \cite{paulydebrecht}. This approach in particular subsumes the join of classical and effective descriptive set theory put forth by \name{Moschovakis} in \cite{moschovakis2}. We will consider uniform versions of two theorems were ordinals appear on one side only: Lusin's separation theorem \cite{lusin} and the Hausdorff-Kuratowski theorem. An effective version of the Lusin separation theorem had already been obtained by \name{Moschovakis} in \cite{moschovakis2}, based on earlier work by \name{Aczel} on Lusin's separation theorem in constructive mathematics \cite{aczel}. The version here is somewhat more general, and the interplay of its proof with the results on computability on the countable ordinals might be illuminating.

\subsection{The computable Lusin separation theorem}
Lusin's separation theorem states that disjoint $\bolds^1_1$-sets in Baire space can be separated by a Borel set. We will split it into two parts: The first has no completeness requirements on the ambient space, makes use of $\min$ (in the form of Proposition \ref{prop:binaryminext}) and provides the separating set. The second requires some form of completeness of the ambient space and translates the separating set into $\mathcal{B}$ as introduced in Subsection \ref{subsec:furtherstructures}.

A very convenient approach to obtain representations of point classes is via characteristic functions into spaces of truth values. We will need to recall the $\Sigma^1_1$ and the Borel truth values, and prove some additional properties of these before we can state and prove our first main result of this section.

\begin{definition}
Let $\mathbb{S}_{\Sigma^1_1}$ be the represented space with underlying set $\{\top,\bot\}$ and representation $\delta_{\Sigma^1_1}$ where $\delta_{\Sigma^1_1}(p) = \top$ iff $p$ codes an ill-founded tree and $\delta_{\Sigma^1_1}(p) = \bot$ iff $p$ codes a well-founded tree.
\end{definition}

We can, by identification of sets and their characteristic functions, view $\mathcal{C}(\mathbf{X},\mathbb{S}_{\Sigma^1_1})$ as the space of $\bolds^1_1$-subsets of $\mathbf{X}$. This is justified by the following:

\begin{observation}
The map $A \mapsto \chi_{\pi_2(A)} : \mathcal{A}(\Baire \times \mathbf{X}) \to \mathcal{C}(\mathbf{X},\mathbb{S}_{\Sigma^1_1})$ is computable and has a computable multivalued inverse.
\end{observation}

The following definition is taken from \cite{pauly-gregoriades}, and refers to the set $\bcode$ of Borel codes from \cite{moschovakisb}, which we recalled in Subsection \ref{subsec:furtherstructures}:

\begin{definition}[(\footnote{This definition differs slightly from the one given in \cite{pauly-gregoriades}, but is equivalent to it.})]
\label{def:boreltruth}
We define the represented space $\mathbf{S}_{\mathcal{B}} = (\{\bot, \top\}, \delta_\mathcal{B})$ recursively via
\begin{align*}
\delta_\mathcal{B}(p) \ \textrm{is defined}& \ \iff \ p \in \bcode\\
\delta_\mathcal{B}(0p) = & \top\\
\delta_\mathcal{B}(1\langle p_0, p_1, \ldots\rangle) = & \bigvee_{i \in \mathbb{N}} \neg \delta_\mathcal{B}(p_i).
\end{align*}
\end{definition}

\begin{proposition}[{\cite[Proposition 42]{pauly-gregoriades}}]
\label{prop:sbcclosure}
The following maps are computable:
\begin{enumerate}
\item $\neg : \mathbb{S}_\mathcal{B} \to \mathbb{S}_\mathcal{B}$
\item $\wedge, \vee : \mathbb{S}_\mathcal{B} \times \mathbb{S}_\mathcal{B} \to \mathbb{S}_\mathcal{B}$
\item $\bigwedge, \bigvee : \mathcal{C}(\mathbb{N},\mathbb{S}_\mathcal{B}) \to \mathbb{S}_\mathcal{B}$
\end{enumerate}
\end{proposition}

We can find even stronger closure properties for $\mathbb{S}_\mathcal{B}$:

\begin{proposition}
\label{prop:forallexistssb}
The following maps are computable:
\begin{enumerate}
\item $\mathalpha{\exists} : \mathcal{C}(\Baire,\mathbb{S}_\mathcal{B}) \to \mathbb{S}_\mathcal{B}$
\item $\mathalpha{\forall} : \mathcal{C}(\Baire,\mathbb{S}_\mathcal{B}) \to \mathbb{S}_\mathcal{B}$
\end{enumerate}
\begin{proof}
Given a function $f \in \mathcal{C}(\Baire,\mathbb{S}_\mathcal{B})$ we can compute a list $(w_i)_{i \in \mathbb{N}}$ of finite prefixes of the input inducing $f$ to output $1$ as the first symbol of the output. We then define functions $f_{ij} \in \mathcal{C}(\Baire,\mathbb{S}_\mathcal{B})$ such that $f(w_ip) = 1\langle f_{i1}(p),f_{i2}(p),\ldots\rangle$. Moreover, we can compute $b_f \in \mathbb{S}$ denoting whether there is some prefix $v$ such that $f(v)$ starts with $0$.

Now we find that $\mathalpha{\exists}(f) = b_f \vee \left ( \bigvee_{i,j \in \mathbb{N}} \mathalpha{\exists}(f_{ij})\right )$ and $\mathalpha{\forall}(f) = b_f \wedge \left ( \bigwedge_{i,j \in \mathbb{N}} \mathalpha{\forall}(f_{ij}) \right )$. The proof that this indeed is a valid algorithm proceeds via induction, entirely analogous to the definition of evaluation trees (Definition \ref{def:ev}) and their relationship to continuous functions (Lemma \ref{lemma:evaluationtree}).
\end{proof}
\end{proposition}

For the following, we need again an extension of $\Cord$ by $\infty$, albeit a tamer one which we shall denote by $\overline{\Cord}_{\textrm{nf}}$. Any $\alpha < \infty$ in $\overline{\Cord}_{\textrm{nf}}$ is represented by a $\delta_{\textrm{nK}}^{\textrm{nf}}$-name. The names for $\infty$ are the least (by inclusion) non-empty set $I \subseteq \Baire$ satisfying that $p \in I$ iff $p = 2\langle p_0,\ldots,\rangle$ where for each $i$, either $p_i \in \dom(\delta_{\textrm{nK}}^{\textrm{nf}})$ or $p_i = 1q$ with $q \in I$, and for at least one $i_0$ \ $p_{i_0} = 1q$ with $q \in I$.

\begin{observation}
$\id : \overline{\Cord}_{\textrm{nf}} \to \overline{\Cord}$ is computable.
\end{observation}

The purpose of $\overline{\Cord}_{\textrm{nf}}$ is that we can compute the rank of a countably branching tree as an element of $\overline{\Cord}_{\textrm{nf}}$, where we understand the rank of an ill-founded tree to be $\infty$.

\begin{proposition}
\label{prop:leq}
$\mathalpha{\leq} : \overline{\Cord}_{\textrm{nf}} \times \Cord \to \mathbf{S}_\mathcal{B}$ is computable.
\begin{proof}
By Proposition \ref{prop:sbcclosure} we may freely use countable boolean operations. We can then compute $\mathalpha{\leq}$ inductively as follows:
\begin{enumerate}
\item $\mathalpha{\leq}(0p,q) = \top$
\item $\mathalpha{\leq}(1p,0q) = \bot$
\item $\mathalpha{\leq}(1p,1q) = \mathalpha{\leq}(p,q)$
\item $\mathalpha{\leq}(1p,2\langle q_0,\ldots, \rangle) = \bigvee_{i \in \mathbb{N}} \mathalpha{\leq}(1p,q_i)$
\item $\mathalpha{\leq}(2\langle p_0,\ldots,\rangle, q) = \bigwedge_{i \in \mathbb{N}} \mathalpha{\leq}(p_i,q)$
\end{enumerate}
The structure of names in $\overline{\Cord}_{\textrm{nf}}$ ensures that any application of Rule 5 has to be followed by some other rule, which in turn allows us to prove correctness by induction over the structure of $q$.
\end{proof}
\end{proposition}

The first component of our version of Lusin's separation theorem will work for spaces which admit an effectively traceable representation. This notion was introduced in \cite{paulybrattka4}, and is further investigated in \cite[Section 7]{paulybrattka4}. Note that in particular any standard representation of an effective topological space (in the sense of Weihrauch \cite{weihrauchd}) is effectively traceable (\cite[Corollary 71]{paulybrattka4}).

\begin{definition}[{\cite[Definition 66]{paulybrattka4}}]
\label{def:efftraceable}
We call a representation $\delta_\mathbf{X} : \subseteq \Baire \to X$ \emph{effectively traceable}, if there is a computable function $T :\subseteq \Baire \times \Baire \to \Baire$ with
$\{T(p,q) : q \in \Baire \} = \delta_\mathbf{X}^{-1}(\delta_\mathbf{X}(p))$ for all $p \in \dom(\delta_\mathbf{X})$ and $\dom(T)=\dom(\delta_\mathbf{X})\times\Baire$.
\end{definition}

Define $\operatorname{Lusin}_\mathbf{X} : \subseteq \mathcal{C}(\mathbf{X},\mathbb{S}_{\Sigma^1_1}) \times \mathcal{C}(\mathbf{X},\mathbb{S}_{\Sigma^1_1}) \to \mathcal{C}(\mathbf{X},\mathbb{S}_{\mathcal{B}})$ by $(\chi_A, \chi_B) \in \dom(\operatorname{Lusin})$ iff $A \cap B = \emptyset$, and $\chi_C \in \operatorname{Lusin}(A,B)$ iff $A \subseteq C$ and $C \cap B = \emptyset$.

\begin{theorem}
Let $\mathbf{X}$ admit an effectively traceable representation $\delta_\mathbf{X}$. Then $\operatorname{Lusin}_\mathbf{X}$ is well-defined and computable.
\begin{proof}
Being given $\chi_A, \chi_B \in \mathcal{C}(\mathbf{X},\mathbb{S}_{\Sigma^1_1})$ means being given realizers $F_A, F_B : \subseteq \Baire \to \Baire$. By computability of the rank of a tree, we can turn these into continuous $f_A, f_B : \subseteq \Baire \to \overline{\Cord}_{\textrm{nf}}$ such that if $\delta_\mathbf{X}(p) \in A$, then $f_A(p) = \infty \in \overline{\Cord}_{\textrm{nf}}$ and of $\delta_\mathbf{X}(p) \notin A$, then $f_A(p) < \infty$, and likewise for $f_B$.

The assumptions on $\chi_A$ and $\chi_B$ imply that the requirements for Proposition \ref{prop:binaryminext} are satisfied for any $p \in \dom(\delta_\mathbf{X})$, and we can thus compute $\min \{f_A(p), f_B(p)\} \in \Cord$ from $p \in \dom(\delta_\mathbf{X})$. Next, we consider $p \mapsto \mathalpha{\leq} (f_B(p), \min \{f_A(p), f_B(p)\}) : \subseteq \Baire \to \mathbb{S}_\mathcal{B}$, which is computable from $f_A$, $f_B$ by Proposition \ref{prop:leq}. Whenever $\delta_\mathbf{X}(p) \in A$, then this is mapped to $\top \in \mathbb{S}_\mathcal{B}$, if $\delta_\mathbf{X}(p) \in A$, the expression returns $\bot$.

However, for $\delta_\mathbf{X}(p) \notin A \cup B$, the resulting truth value might differ for different names of the same point. We thus take the computable witness $T$ that $\delta_\mathbf{X}$ is effectively traceable, and combine this with $\mathalpha{\forall}$ from Proposition \ref{prop:forallexistssb} to:
\[f_C := p \mapsto \mathalpha{\forall} \left ( q \mapsto \mathalpha{\leq} (f_B(T(p,q)), \min \{f_A(T(p,q)), f_B(T(p,q))\}) \right )\]
Finally, note that $f_C \circ \delta_\mathbf{X}^{-1} \in \operatorname{Lusin}(\chi_A,\chi_B)$.
\end{proof}
\end{theorem}

The following was already shown as \cite[Theorem 41]{pauly-gregoriades}. However, the proof there invoked the Suslin-Kleene theorem, which we wish to obtain as a corollary here.

\begin{theorem}
Let $\mathbf{X}$ be quasi-Polish. Then $\chi_A \mapsto A : \mathcal{C}(\mathbf{X},\mathbb{S}_\mathcal{B}) \to \mathcal{B}(\mathbf{X})$ is well-defined and computable.
\begin{proof}
Recall from \cite{debrecht6} that the quasi-Polish spaces $\mathbf{X}$ are exactly those represented spaces admitting a total effectively open representation $\delta_\mathbf{X} : \Baire \to \mathbf{X}$. We again employ a similar technique to the one via Definition \ref{def:ev} and Lemma \ref{lemma:evaluationtree}), which we already adapted for Proposition \ref{prop:forallexistssb}, and proceed by induction over the corresponding analogue of the evaluation tree of $\chi_A$.

We monitor the execution of a realizer $F$ of $\chi_A$ to obtain a list of prefixes $(w_i)_{i \in \mathbb{N}}$ causing the prefix $1$ of the output, and prefixes $(v_i)_{i \in \mathbb{N}}$ causing the prefix $0$.

We can conclude immediately that any $x \in \mathbf{X}$ with some $\delta_\mathbf{X}$-name $v_ip$ lies in $A$. As $\delta_\mathbf{X}$ is effectively open, we can compute $A_0 := \delta_{\mathbf{X}}[\bigcup_{i \in \mathbb{N}} v_i\Baire] \in \mathcal{O}(\mathbf{X})$.

For any $w_i$, we consider the maps $f_{ij}$ such that $F(w_ip) = 1\langle f_{i1}(p),f_{i2}(p),\ldots\rangle$. We recursively apply the algorithm to the $f_{ij}$ (ignoring the minor issue that $f_{ij}$ technically is not guaranteed to have the correct type) to obtain sets $A_{ij}$ such that $x \in A_{ij}$ iff $x$ has some name $w_ip$ with $\delta_\mathcal{B}(f_{ij}(p)) = \top$.

Finally, we find that $A = A_0 \cup \left (\bigcup_{i \in \mathbb{N}} \bigcap_{j \in \mathbb{N}} A_{ij}^C \right )$.
\end{proof}
\end{theorem}

We can further conclude that the Suslin-Kleene theorem holds in quasi-Polish spaces, generalizing previous results by \name{Moschovakis} \cite[Section 7B]{moschovakisb} and \name{Selivanov} \cite[Theorem 4]{selivanov8}:

\begin{corollary}
For any quasi-Polish space $\mathbf{X}$ there is an effective procedure to compute a Borel code of some set $C$ from analytic codes of disjoint sets $A$ and $B$ such that $A \subseteq C$ and $C \cap B = \emptyset$.
\end{corollary}

\begin{corollary}
In quasi-Polish spaces, the $\Delta^1_1$-sets coincide uniformly with the Borel sets.
\end{corollary}

\subsection{The computable Hausdorff-Kuratowski theorem}
\label{subsec:hk}
We shall now prepare the formulation of the Hausdorff-Kuratowski theorem in the framework of computable endofunctors on the category of represented spaces as introduced by \name{de Brecht} and \name{P.} in \cite{paulydebrecht2,pauly-descriptive-lics,debrecht5}. The setting closely follows the corresponding section in \cite{debrecht5} by \name{de Brecht}, where a weaker (and non-effective) version of our desired result was proven\footnote{The version of the Hausdorff-Kuratowski theorem from \cite{debrecht5} was also generalized in \cite{becher}, albeit still only in non-effective ways.}.

As preparation for the formulation and proof of a computable Hausdorff-Kuratowski theorem, we introduce a tailor-made representation of the countable ordinals. Let a \emph{nice relation} be a well-founded quasi-order $\preceq$ on $\mathbb{N}$, such that $\forall n, \ n \preceq 0$, and whenever $n \prec m$, then $n > m$.

\begin{definition}
\label{def:nice}
We define a representation $\delta_{\textrm{nR}} : \subseteq \Cantor \to \cord$ by $\delta_{\textrm{nR}}(p) = \alpha$, iff the relation $\preceq_{p}$ defined via $n \preceq_{p} m$ iff $p(\langle n, m\rangle) = 1$ is a nice relation of height $\alpha + 1$ (the height of any nice relation is a countable successor ordinal, and every countable successor ordinal occurs as the height of a nice relation).
\end{definition}

\begin{proposition}
$\id : \Cord \to (\cord, \delta_{\textrm{nR}})$ is a computable isomorphism.
\begin{proof}
It is straight-forward to verify that $0$, $\mathalpha{+1}$, $\sup_{i \in \mathbb{N}}$ and $\operatorname{Lower}$ are computable operations on $(\cord,\delta_\textrm{nR})$. The claim then follows from Theorem \ref{theo:charac}.
\end{proof}
\end{proposition}

For any sequence of countable ordinals $(\alpha_i)_{i \in \mathbb{N}}$, we define a function $L_{(\alpha_i)_{i \in \mathbb{N}}} : \subseteq \Baire \to \Baire$. The sequence only impacts the domain, but whenever $L_{(\alpha_i)_{i \in \mathbb{N}}}(p)$ is defined, then $2L_{(\alpha_i)_{i \in \mathbb{N}}}(p)(n) = p(\max \{i \in \mathbb{N} \mid p(i) \textnormal{ is odd}\} + n + 1)$; i.e.~$L_{(\alpha_i)_{i \in \mathbb{N}}}$ takes the maximal tail of its input consisting of only even values, and returns the result of pointwise division by $2$. Obviously any sequence in the domain of $L_{(\alpha_i)_{i \in \mathbb{N}}}$ has to contain only finitely many odd entries; and we additionally demand that for $p \in \dom(L_{(\alpha_i)_{i \in \mathbb{N}}})$, if $n < m$, and $p(n) = 2k+1$ and $p(m) = 2j+1$, then $\alpha_k > \alpha_j$.

\begin{definition}
We define a computable endofunctor $\mathfrak{L}_{(\alpha_n)_{n \in \mathbb{N}}}$ by $\mathfrak{L}_{(\alpha_n)_{n \in \mathbb{N}}}\left (X, \delta \right ) = (X, \delta \circ L_{(\alpha_i)_{i \in \mathbb{N}}})$ and the straightforward extension to functions.
\end{definition}

Each endofunctor $\mathfrak{L}_{(\alpha_n)_{n \in \mathbb{N}}}$ captures a version of computability with finitely many mindchanges (e.g.~\cite{ziegler2,ziegler3}): The regular outputs are encoded as even numbers. Finitely many times, the output can be reset by using an odd number, however, when doing so, one has to count down within the list of ordinals parameterizing the function (which in particular ensures that it happens only finitely many times). We thus find it connected to the \emph{level} introduced by \name{Hertling} \cite{hertling}, and further studied by him and others in \cite{hertling3,hertling4,hertling8,paulymaster,paulyreducibilitylattice,debrecht5}.

\begin{definition}
Given a function $f : \subseteq \Baire \to \Baire$, we define the sets $\mathcal{L}_\alpha(f) \subseteq \Baire$ inductively via:
\begin{enumerate}
\item $\mathcal{L}_0(f) = \dom(f)$
\item $\mathcal{L}_{\alpha + 1}(f) = \overline{\{x \in \mathcal{L}_{\alpha}(f) \mid f|_{\mathcal{L}_{\alpha}} \textnormal{ is discontinuous at } x\}}$
\item $\mathcal{L}_{\gamma}(f) = \bigcap_{\beta < \gamma} \mathcal{L}_\beta(f)$ for limit ordinals $\gamma$.
\end{enumerate}
Then we say $\Lev(f) := \min \{\alpha \mid \mathcal{L}_\alpha(f) = \emptyset\}$.
\end{definition}

\begin{theorem}
\label{theo:lev}
If $f : \Baire \to \mathfrak{L}_{(\alpha_i)_{i \in \mathbb{N}}}\Baire$ is continuous, then $\Lev(f) \leq \left ( \sup_{i \in \mathbb{N}} \alpha_i \right ) + 1$.
\begin{proof}
Let $F$ be a continuous realizer of $f$. For any $n \in \mathbb{N}$, the set $U_n := \{p \in \Baire \mid \exists k \ F(p)(k) = 2n+1\}$ is open. Let $n$ be such that $\alpha_n$ is the smallest ordinal in $(\alpha_i)_{i \in \mathbb{N}}$. Then $f|_{U_n}$ is continuous, as there cannot be any further mindchanges happening, i.e.~$\mathcal{L}_1(f) \subseteq U_n^C$. Then consider $m$ such that $\alpha_m$ is the second smallest ordinal in $(\alpha_i)_{i \in \mathbb{N}}$, and conclude $\mathcal{L}_2(f) \subseteq \left (U_n \cup U_m \right)^C$. Iterating this process yields $\mathcal{L}_{\sup_{i \in \mathbb{N}} \alpha_i}(f) \subseteq \left (\bigcup_{i \in \mathbb{N}} U_i \right )^C$, and we notice that $f|_{\left (\bigcup_{i \in \mathbb{N}} U_i \right )^C}$ does not make any mindchanges, hence is continuous. Thus $\mathcal{L}_{\left (\sup_{i \in \mathbb{N}} \alpha_i\right ) +1}(f) = \emptyset$.
\end{proof}
\end{theorem}

\begin{proposition}
Let $(\alpha_i)_{i \in \mathbb{N}}$ be such that $\exists \alpha \in \cord$ with $\{\alpha_i \mid i \in \mathbb{N}\} = \{\beta \in \cord \mid \beta < \alpha\}$. Then $\Lev(L_{(\alpha_i)_{i \in \mathbb{N}}}) = \alpha + 1$.
\begin{proof}
(Sketch): In this situation, the set inclusions in the proof of Theorem \ref{theo:lev} are tight.
\end{proof}
\end{proposition}

The computable Hausdorff-Kuratowski theorem has at its heart a dependent sum type; namely the construction $\sum_{(\alpha_i)_{i \in \mathbb{N}} \in \Cord^\mathbb{N}} \left (\mathcal{C}(\mathbf{X},\mathfrak{L}_{(\alpha_i)_{i \in \mathbb{N}}}\mathbf{Y}) \right )$ for some represented spaces $\mathbf{X}$, $\mathbf{Y}$. A point in this space is a pair, consisting of a sequence of countable ordinals and a function $f : \mathbf{X} \to \mathbf{Y}$, the latter given only in a $\mathfrak{L}_{(\alpha_i)_{i \in \mathbb{N}}}$-continuous way.

\begin{theorem}[Computable Hausdorff-Kuratowski theorem]
\label{theo:hk}
Let $\mathbf{X}$, $\mathbf{Y}$ be represented spaces, and $\mathbf{X}$ be complete. Then the map $\textrm{HK} : \mathcal{C}(\mathbf{X},\mathbf{Y}^\nabla) \mto \sum_{(\alpha_i)_{i \in \mathbb{N}} \in \Cord^\mathbb{N}} \left (\mathcal{C}(\mathbf{X},\mathfrak{L}_{(\alpha_i)_{i \in \mathbb{N}}}\mathbf{Y}) \right )$ where $((\alpha_i)_{i \in \mathbb{N}}, g) \in \textrm{HK}(f)$ iff $f = g$, is computable.
\begin{proof}
The general case reduces to the situation where $\mathbf{X} = \mathbf{Y} = \Baire$: As a complete represented space, $\mathbf{X}$ has a total representation $\delta_\mathbf{X} : \Baire \to \mathbf{X}$. We can then operate on a realizer of the original $f$, as all involved endofunctors are derived from jump operators.

That we have $f \in \mathcal{C}(\Baire,(\Baire)^\nabla)$ means we may evaluate $f$ with finitely many mindchanges. Any such mindchange occurs after a finite prefix of the input has been read. Thus, we may identify countably many \emph{mindchange occurrences}. Using $0 \in \mathbb{N}$ to denote no mindchange at all, we can proceed to obtain a relation $\preceq$ and a numbering of the mindchange occurrences, such that if mindchange $n$ occurs after mindchange $m$, then $n \prec m$. If we are not aware of any not-yet-numbered mindchange occurrences, we just allocate the next natural number to the \emph{none-mindchange} at $0$ again.

We adjust the realizer for $f$ in a way such that any regular output $n$ is replaced by $2n$, and a mindchange symbol corresponding to the $m$-th mindchange is replaced by $2m+1$.

As $f$ is total, we find that any decreasing chain through $(\mathbb{N}, \prec)$ corresponds to the mindchanges made for some input to $f$. Thus, the relation $\prec$ is well-founded, and the other properties of a nice relation (cf.~Definition \ref{def:nice}) follow from the construction. Using the operation $\operatorname{Lower}$, we can identify for each $n \in \mathbb{N}$ the corresponding ordinal of its height in the relation, yielding the sequence $(\alpha_i)_{i \in \mathbb{N}}$.
\end{proof}
\end{theorem}

\begin{corollary}
Let $f : \mathbf{X} \to \mathbf{Y}$ be computable with finitely many mindchanges, and $\mathbf{X}$ be complete. Then $\textrm{Lev}(f)$ exists and is a computable ordinal.
\begin{proof}
Combine Theorem \ref{theo:hk} with Theorem \ref{theo:lev}.
\end{proof}
\end{corollary}

The result of the preceding corollary was also announced by \name{Selivanov} at CCA 2014.

\section{Iteration over some ordinal}
\label{sec:iteration}
Having a suitable notion of computability of the countable ordinals available allows us to define an important notion in the study of Weihrauch reducibility: What it means to iterate some (potentially non-computable) principle over some given countable ordinal. There are some similarities to how $\textrm{ATR}_0$ is used in reverse mathematics -- this principle is often informally characterized as being able to iterate the Turing jump over some countable ordinal, cf.~\cite{greenberg}. A notable difference is that in reverse mathematics, one has to prove that the ordinal one wants to iterate over is indeed an ordinal within the confines of the given system, whereas in the Weihrauch lattice, we only need to construct the ordinal effectively, but can use classical logic to prove well-foundedness.

For simplicity, we first define ordinal iteration only for functions on Baire space, followed by the more complicated definition for multivalued functions. As any $f : \subseteq \mathbf{X} \mto \mathbf{Y}$ is Weihrauch-equivalent to $\delta_\mathbf{Y} \circ f \circ \delta_\mathbf{X}^{-1} : \subseteq \Baire \mto \Baire$, it is sufficient to give the definition for those.

\begin{definition}[Singlevalued case]
Fix a standard enumeration $(\Phi_n)_{n \in \mathbb{N}}$ of the computable functions $\Phi_n : \subseteq \Baire \to \Baire$. Given a partial function $f : \subseteq \Baire \to \Baire$, we define $f^\dag : \subseteq \Baire \to \Baire$ as follows:
\begin{enumerate}
\item $f^\dag(\langle 0p,q\rangle) = q$
\item $f^\dag(\langle 1p,nq\rangle) = \langle f(\Phi_{n}(f^\dag(\langle p, q\rangle))), f^\dag(\langle p, q\rangle)\rangle$
\item $f^\dag(\langle 2\langle p_0,p_1,\ldots\rangle, \langle q_0,q_1,\ldots\rangle\rangle) = \langle f^\dag(\langle p_0,q_0\rangle), f^\dag(\langle p_1,q_1\rangle), \ldots \rangle$
\end{enumerate}
\end{definition}

\begin{definition}[General case]
Fix a standard enumeration $(\Phi_n)_{n \in \mathbb{N}}$ of the computable functions $\Phi_n : \subseteq \Baire \to \Baire$. Given a partial multivalued function $f : \subseteq \Baire \to \Baire$, we define $f^\dag : \subseteq \Baire \mto \Baire$ as follows:
\begin{enumerate}
\item $f^\dag(\langle 0p,q\rangle) = q$
\item $r \in f^\dag(\langle 1p,nq\rangle)$ iff $\exists h \quad r \in \langle f(\Phi_{n}(h)), h\rangle \wedge h \in f^\dag(\langle p, q\rangle)$
\item $f^\dag(\langle 2\langle p_0,p_1,\ldots\rangle, \langle q_0,q_1,\ldots\rangle\rangle) = \langle f^\dag(\langle p_0,q_0\rangle), f^\dag(\langle p_1,q_1\rangle), \ldots \rangle$
\end{enumerate}
\end{definition}

Essentially, the first argument of $f^\dag$ is a $\delta_{\textrm{nK}}$-name for some ordinal (that is iterated over); while the second argument provides the actual inputs to the function $f$. In particular, $\dom(f^\dag) \subseteq \langle \dom(\delta_{\textrm{nK}}), \Baire \rangle$. Thus, we can define $f^{\dag,\alpha}$ for some ordinal $\alpha$ as the restriction of $f^\dag$ to $\langle \delta_{\textrm{nK}}^{-1}(\{\beta \mid \beta < \alpha\}), \Baire\rangle$. Thus, trivially, if $\beta \leq \alpha$, then $f^{\dag,\beta} \leqW f^{\dag,\alpha}$. Additionally, we point out that $f^{\dag,0} \equivW \emptyset$, $f^{\dag,1} \equivW \id$, $f^{\dag,2} \equivW \widehat{(f + \id)}$ and more generally, $f^{\dag,n+1} \equivW \widehat{(f + \id)}^{(n)}$.

We list some further simple properties of $^\dag$:

\begin{proposition}
\begin{enumerate}
\item If $f \leqW g$, then $f^\dag \leqW g^\dag$ and $f^{\dag,\alpha} \leqW g^{\dag,\alpha}$.
\item $\widehat{f^\dag} \equivW f^\dag$
\item $f \star f^\dag \equivW f^\dag$
\item If $\widehat{(f + \id)} \star \widehat{(f + \id)} \equivW f$, then $f^\dag \equivW f$.
\end{enumerate}
\begin{proof}
\begin{enumerate}
\item Straight-forward.
\item Essentially, this is just replacing $\alpha \in \Cord$ by $\sup_{i \in \mathbb{N}} \alpha$.
\item Essentially, this is just replacing $\alpha \in \Cord$ by $\alpha + 1$, together with the definition of $\star$ from \cite{paulybrattka4}.
\item By induction over the first parameter. The first case uses $\id \leqW f$, the second case uses $f \equivW f \star f$, and the third case uses $f \equivW f \star \widehat{f}$.
\end{enumerate}
\end{proof}
\end{proposition}

Unlike other unitary operations introduced on the Weihrauch lattice such as $\widehat{\phantom{a}}$ \cite{brattka2}, $^*$ \cite{paulyreducibilitylattice} and $^\diamond$ \cite{paulyneumann}, we find that $^\dag$ is not a closure operator. The reason for this is that if we can obtain some non-computable ordinals with help of $f$, then we can recurse much longer using $(f^\dag)^\dag$ than we could just using $f^\dag$. In particular, not even $f^\dag \star f \leqW f^\dag$ holds in general.

With the help of $\dag$, we can characterize the principle $\UC_\Baire : \subseteq \mathcal{A}(\Baire) \to \Baire$, which is the restriction of $\C_\Baire : \subseteq \mathcal{A}(\Baire) \mto \Baire$ (as mentioned in Subsection \ref{subsec:borel}) to singletons. We will express $\UC_\Baire$ in terms of $\lpo : \Cantor \to \{0,1\}$ mapping $0^\mathbb{N}$ to $1$ and $p \neq 0^\mathbb{N}$ to $0$, as well as $\lim : \subseteq \Baire \to \Baire$ defined via $\lim(p)(n) = \lim_{i \to \infty} p(\langle n,i\rangle)$. All these principles are discussed in \cite{paulybrattka}. In \cite{brattka}, the degree of $\lim^{(n)}$ is shown to be complete for $\Sigma^0_{n+1}$-measurable functions.

Before we can prove the theorem, we will formulate one ingredient separately, as it may be of independent interest. Namely, we show that using $\lpo^\dag$ we can translate Borel truth values into the ordinary booleans $\mathbf{2}$ (with the discrete topology).
\begin{lemma}
\label{lem:boreltoregtruth}
$\left ( \id : \mathbb{S}_\mathcal{B} \to \mathbf{2} \right ) \leqW \lpo^\dag$
\begin{proof}
The map $(b_i)_{i \in \mathbb{N}} \mapsto \bigvee_{i \in \mathbb{N}} \neg b_i : \mathcal{C}(\mathbb{N},\mathbf{2}) \to \mathbf{2}$ is Weihrauch equivalent to $\lpo$. Thus, $\lpo^\dag$ suffices to follow the assignment of truth values in $\mathbb{S}_\mathcal{B}$ (Definition \ref{def:boreltruth}).
\end{proof}
\end{lemma}

\begin{theorem}
\label{theo:lpodag}
$\lpo^\dag \equivW \lim^\dag \equivW \UC_\Baire$.
\begin{proof}
As $\widehat{\lpo} \equivW \lim$ \cite{brattka2}, it follows that $\lpo^\dag \equivW \lim^\dag$. We proceed to prove that $\lpo^\dag \leqW \UC_\Baire$. For this, we need to construct some $\{r\} \in \mathcal{A}(\Baire)$ from $\langle p, q\rangle \in \dom(\lpo^\dag)$ such that from $r$ we can compute $\lpo^\dag(\langle p, q\rangle)$. By currying, it suffices to show that given $p$, $q$ and a candidate for $r$, we can reject unsuitable candidates. We thus proceed as follows:
\begin{enumerate}
\item If $p = 0p'$, then reject iff $r \neq \langle q, 0^\mathbb{N}\rangle$.
\item If $p = 1p'$ and $q = nq'$, split $r = \langle \langle r_1, r_2\rangle,kh\rangle$. Recursively work on $p$, $q'$ and $\langle r_2,h\rangle$. Simultaneously, apply $\Phi_n$ to $r_2$. If $r_1 = 1$, reject if $k \neq 0$ or if $\Phi_n$ ever outputs a number but $0$. If $r_1 = 0$, reject if $\Phi_n$ does not output the first $1$ after exactly $k$ steps.
\item If $p = 2\langle p_1,p_2,\ldots\rangle$ and $q = \langle q_1,q_2,\ldots\rangle$, split $r = \langle \langle r_1,r_2,\ldots\rangle, \langle h_1,h_2,\ldots\rangle\rangle$. Now recursively work on each triple $p_i$, $q_i$, $\langle r_i,h_i\rangle$.
\end{enumerate}

We can verify that for any $\langle p, q\rangle \in \dom(\lpo^\dag)$ there is exactly one $r = \langle r',h\rangle$ which will not be rejected at some stage. Moreover, for this unique $r$ we find that $r' = \lpo^\dag(\langle p,q\rangle)$, hence, we have completed the proof of $\lpo^\dag \leqW \UC_\Baire$.

Now let us proceed to the direction $\UC_\Baire \leqW \lpo^\dag$. As closed subsets of $\Baire$ can be represented as sets of infinite branches through some tree, $\UC_\Baire$ equivalently is the task of finding the infinite path through a countably-branching tree $T$ having exactly one infinite branch. For any $w \in \mathbb{N}^*$, let $T_w := T \setminus w\{v \in \mathbb{N}^*\}$. By assumption on $T$, for any $n$ there is exactly one $w \in \mathbb{N}^n$ such that $T_w$ is well-founded. Using computability of the rank of a tree from Subsection \ref{subsec:furtherstructures} together with $\min$ from Proposition \ref{prop:minext}, we can compute $\alpha_w \in \overline{\Cord}_{\textrm{nf}}$, the maximum of the height of $T_w$ and $\omega$, and then $\alpha_n \in \Cord$ as $\alpha_n := \min_{w \in \mathbb{N}^n} T_w$. Now we find that $\alpha_w = \alpha_n$ iff $w$ is the unique vertex at level $n$ that lies on the unique infinite branch. Let $b_w := \mathalpha{\leq}(\alpha_w,\alpha_n) \in \mathbb{S}_\mathcal{B}$ be obtained via Proposition \ref{prop:leq}. We then use Lemma \ref{lem:boreltoregtruth} to find $b_w \in \mathbf{2}$ with the help of $\widehat{\lpo^\dag} \equivW \lpo^\dag$. Reconstructing the unique path from the $b_w \in \mathbf{2}$ is then straight-forward, as we can decide whether some $w \in \mathbb{N}^*$ is a valid prefix.
\end{proof}
\end{theorem}

Iterated application of Shoenfield's limit lemma shows that any iterated Turing jump $\emptyset^{(\alpha)}$ for some computable ordinal $\alpha$, and subsequently any hyperarithmetical degree, appears as the output of $\lim^\dag$ on some computable input. Combining Theorem \ref{theo:lpodag} with Theorem \ref{theo:cbaireremoval} then yields the following corollary, which duplicates a well-known theorem by Spector \cite{spector}:

\begin{corollary}
If $\alpha \in \Cord$ is hyperarithmetical, then $\alpha$ is already computable.
\end{corollary}

As a further application of Theorem \ref{theo:lpodag} and the computable Hausdorff-Kuratowski theorem (Theorem \ref{theo:hk}), we point out that they can be combined with the results from \cite{paulyleroux3-arxiv} to conclude that $\Delta^0_2$-determinacy over Cantor space is Weihrauch equivalent to $\lpo^\dag$.

\bibliographystyle{eptcs}
\bibliography{../../spieltheorie}

\section*{Acknowledgements}
I am grateful to Victor Selivanov for sparking my interest in a computable version of the Hausdorff Kuratowski theorem and to Vasco Brattka and Matthew de Brecht for various discussions on this question. The comparison of the various representations of the countable ordinals started with a discussion with Vassilios Gregoriades. A remark by Takayuki Kihara was crucial for the formulation of Theorem \ref{theo:lpodag}. I am indebted to Alberto Marcone and Andrea Cetollo for pointing out a mistake in an earlier version of Theorem \ref{theo:newsup}; and to Hugo Nobrega for helpful discussion.

This work benefited from the Royal Society International Exchange Grant IE111233 and the Marie Curie International Research Staff Exchange Scheme \emph{Computable Analysis}, PIRSES-GA-2011- 294962. The author was partially supported by the ERC inVEST (279499) project.

\end{document}